\let\latexaddtocontents\addtocontents

\documentclass[a4paper,twocolumn,11pt]{quantumarticle}
\pdfoutput=1
\usepackage{amsmath, amssymb, amsthm}
\usepackage{graphicx}
\usepackage{hyperref}
\hypersetup{colorlinks=true, citecolor=magenta, linkcolor=blue,urlcolor=teal}

\let\addtocontents\latexaddtocontents

\usepackage[backend=bibtex,doi=true]{biblatex} 
\title{Quantum limit cycles from phase coupling: A framework for NISQ thermodynamic devices}

\author{Nana Siddhartha}
\date{12 October 2026}

\affiliation{Institute of Physics, Slovak Academy of Sciences, Dúbravská cesta 9, 845 11 Bratislava, Slovakia}
\affiliation{Department of Theoretical Physics, Comenius University, Bratislava,Slovakia}
\affiliation{Chennai Mathematical Institute, SIPCOT IT Park, Siruseri 603103, India}

\begin{document}
\begin{abstract}
Quantum decoherence does not permit the operation of (new) miniaturised devices or protocols that can otherwise run unassisted in classical settings, without explicit user agent intervention in the quantum realm. Classical driving signals are inadequate to model a truly quantum device built entirely in the quantum realm. Synchronisation endures miniaturisation without sacrificing quantum coherence, unlike thermalisation. We quantise the Kuramoto model and the Van der Pol oscillator and couple it with qubit(s) to show a robust limit cycle even in typical quantum thermodynamical settings. Novel effects such as Entanglement State Revivals, Zeno locking and thermal current interference are observed. We propose toy model devices that can be implemented harnessing the Synchronisation of a qubit (or an arbitrary collection of qubits) with a nonlinear phase oscillator.  The framework theory used here, Spectral Mechanics, is a fresh look at the foundations of conventional quantum mechanics and is extensible to completely new forms of interactions and models in experimental settings.
\end{abstract}


\maketitle

Huygens first characterised the simultaneous swinging of pendulums sharing a common support as a collective phenomenon of ``sympathy'' \cite{Huygens1673}. This historical observation laid the groundwork for our modern understanding of classical phase locking and frequency entrainment in coupled dynamical systems \cite{Pikovsky2001}. In recent decades, this expanded the same structural principles to demonstrate how spontaneous coordination maps across vast, complex networks in nature \cite{Strogatz2003}. The same tracking of rhythms has since been observed in a wide variety of complex biological architectures. For instance, collective locking acts as a crucial neural signalling protocol during cortical bursting episodes \cite{Amitai1989},  and biological clocks underpinning individual cellular rhythms in \cite{Winfree1967, Winfree2001}. Similar autonomous feedback loops govern the limit-cycle behaviour of metabolic networks and macro-cellular cardiac cycles \cite{Goldbeter1996}. At the microscopic limit, Yoshiki Kuramoto abstracted these collective dynamics into an all-to-all sinusoidal mean-field formulation, defining the canonical threshold for spontaneous phase transitions \cite{Kuramoto1984}. Beyond biological domains, phase locking serves as a feedback mechanism that can lock periodic and even chaotic motion in engineered electronic circuits \cite{Pecora1990}. Historical treatments confirmed these stabilisation thresholds, providing structural guidelines for triode oscillators and modern frequency tuning systems \cite{Adler1946, Appleton1922}. Related collective phase structures appear natively in high-frequency solid-state networks, including macroscopic Josephson junction arrays \cite{Blekhman1988}, as well as coupled laser cavities and macro-acoustic arrays \cite{Rayleigh1896}.

Experimental observations in the early twentieth century led to the van der Pol oscillator as the defining model with which to study non-linear Synchronisation thresholds \cite{vanderPol1926, vanderPol1928}. Structurally, it modifies a simple harmonic oscillator with a non-linear friction term that injects energy at small displacement amplitudes and dissipates it at large amplitudes, thereby selecting a unique, attracting limit-cycle orbit in phase space. Non-conservative models that exhibit these spontaneous orbits share a distinct geometric hallmark: their asymptotic motion settles onto stable periodic trajectories rather than static foci, manifesting a robust structural resistance to external phase perturbations \cite{Pikovsky2001, Kuramoto1984}. This stabilising metric ensures that self-sustained attractors can dynamically modulate their natural frequencies to achieve precise entrainment with coupled partners \cite{Adler1946}. This self-correcting alignment endures even when an external agent subsequently forces a displacement in the underlying state trajectory. Experiments have validated this non-linear phenomenology across a wide span of scales, ranging from molecular assemblies up to mesoscopic and macroscopic length scales \cite{Pikovsky2001, Strogatz2003, Winfree2001}. Naturally, a central open question is the rigorous extension of these attractors into quantum analogues \cite{Zhirov2006, Zhirov2009}. Recent interest in quantum hardware devices—such as trapped-ion crystals, superconducting circuit networks, and optomechanical cavities—makes the systematic investigation of these non-linear structures essential for engineering noise-resilient hardware architectures \cite{Lee2013, Lee2014}. Understanding how these models perform under steady-state thermal currents and non-equilibrium dissipation is vital for scaling modern quantum hardware systems \cite{Walter2014, Walter2015}. Furthermore, these phase-locking architectures open alternative pathways for mitigating macro-environmental decoherence channels while characterising the thermodynamic limits of complex, multi-component computing substrates \cite{Sonar2018, Vinjanampathy2016, Goold2016}.

A priori, we might assume that phase Synchronisation is a fragile quantum state that can be easily dissolved by contact with classical measurement devices or a standard thermal bath. Synchronisation empirically persists well into the macroscopic domain even for open quantum systems that support a stable limit cycle \cite{Lee2013, Walter2014}. Quantum systems driven by these synchronised attractors can be harnessed to operate as high-precision sensors immune to statistical fluctuations \cite{Dutta2019}. In the presence of quantum fluctuations, strict frequency locking is replaced by a smooth crossover from weak to strong phase entrainment regimes \cite{Walter2014, Walter2015}. Interestingly, pure energy quantisation can act as a structural block, preventing identical quantum phase systems from locking together \cite{Lorch2017, Ishibashi2017}. Nonetheless, the non-perturbative locked cycle survives, and its structural persistence permits the creation of novel heat devices that one might think impossible. These include high-efficiency quantum interferometers and thermal-current amplifiers \cite{Jaseem2020}. Such architectures function as the non-linear analogs of standard circuit components previously restricted to driven-dissipative Markovian contexts, such as asymmetric subharmonic three-terminal quantum transistors \cite{Scovil1959} and directional thermal diodes \cite{Kosloff2013_AnnuRev}. These solid-state heat engines provide precise instrumentation mechanisms for regulating quantum energy currents at the nanoscale \cite{Mitchison2021, Pekola2015}.

Implementing a complex quantum circuit will require a mitigating mechanism to attenuate thermal fluctuations in addition to logical mechanisms to protect from errors. A simple solution would be to cool down the substrate on which the computation happens. Anticipating which operations will dissipate more heat and which parts require cooling will require precise control over the dissipation and need a new understanding of non-linear phenomena. It becomes essential to study the flow of heat and its interactions in that context. Obviously, one may use Synchronisation as a natural anharmonic extension to linear signals i.e. as the most easily available interaction at the next order to linear coupling between harmonic oscillators.

Synchronisation introduces several unique features that are highly advantageous for countering decoherence during the operation of an active quantum gate or protocol \cite{Giorgi2012, Manzano2013}. Synchronised systems display phase entrainment while remaining locked onto a stable attractor shell; consequently, brief interactions with an external control agent do not destroy the underlying phase-locking loop \cite{Galve2017, EsHaqiSani2020}. Instead, the internal frequencies shift dynamically to accommodate the perturbation, behaving entirely differently from standard linear systems that rely on rigid normal modes \cite{Pikovsky2001, Strogatz2003}. In conventional architectures, normal modes represent a fragile collective configuration of coupled linear oscillators that easily mixes with environmental noise, radiates energy, and decays into a passive equilibrium state. Thermal equilibrium is destructive to quantum superpositions and is widely identified with decoherence and a separable state \cite{BreuerPetruccione, Rivas2014, Alicki2007}. A state devoid of phase coherence is of limited use for quantum tasks. Entrainment and phase coherence are instead dynamical features of a driven-dissipative attractor and need not suffer the same mixing \cite{Jaseem2020, Sonar2018}. Because these states sit on a stable non-conservative backbone, phase-locked quantum arrays maintain a non-vanishing coherence loop that is structurally protected from local thermalisation channels \cite{Roulet2018, Bastidas2015}.

Unfortunately, such non-conservative dynamics cannot be accommodated via stereotypical completely positive (CP) maps or standard quantum channels \cite{Lindblad1976, Gorini1976, Wolf2008}. While a running system can occasionally be approximated by a one-parameter family of such operations, the partial trace operation forces the reduced density matrix to evolve strictly as a contractive map on the state space \cite{Spohn1976, Rivas2014}. The asymptotic configuration of such operations is strictly bounded to a stationary fixed point, ruling out the generation of a limit cycle with a finite periodic orbit. Equivalently, the standard Lindblad superoperator cannot possess purely imaginary eigenvalues; every oscillatory configuration is systematically damped over time \cite{Lindblad1976, Gorini1976, BreuerPetruccione}. This damping forces the system's coherences to decay to zero, eliminating any persistent, self-sustained dynamics \cite{Albert2014, Carmichael1993, Gardiner2004}. Genuine limit-cycle motion therefore requires additional structure—nonlinear jump operators, time-periodic driving, or an explicit phase degree of freedom—beyond a standard Markovian generator \cite{Lee2013, Walter2014}. 

Our approach is fundamentally different. Here, the oscillator remains completely autonomous and utilises a natural, sinusoidal phase interaction. Instead of manually introducing an artificial dissipator, the stabilising gain is derived directly from the underlying influence functional of the Kuramoto or Van der Pol kernels. This explicit comparison clarifies our unique mechanism. To track these non-Markovian dynamics accurately, alternative operational actions must be deployed to capture the full non-perturbative phase tracking of the quantum oscillator environment \cite{Dutta2019, Klaers2017}.

It makes sense to look for new formalisms that can accommodate such ``nonlinear'' systems when they interact with a quantum degree of freedom \cite{Jaseem2020,Binder2018}. This paper provides such a theory and subsequently examines the novel mesoscale devices that can be constructed if phase oscillators interact directly with a typical finite-dimensional quantum system.

By exploring these structural interactions, we show that the presence of synchronised attractors allows for alternative non-equilibrium quantum thermodynamic platforms. The effect of thermal fluctuations or heat currents on a synchronised limit cycle can be analysed comprehensively only within this unique framework. Such an analysis is crucial to the effort of creating robust quantum devices that can be structurally protected from local environment noise, or even thermally accelerated, rather than destroyed by decoherence . Ultimately, understanding these non-conservative path histories maps out an actionable blueprint for engineering nanoscale energy transport systems in open environments anticipated in \cite{Vinjanampathy2016,Binder2018}.

\subsection*{Outline}
Section \ref{sec:theory} describes the new theory that enables quantisation of phase oscillators. Section \ref{sec:kuramoto}  and \ref{sec:vdp} contain the derivation of the kernels for the Kuramoto and Van der Pol  oscillators respectively. Section \ref{sec:results} describes the coupling of the phase oscillator with a finite-dimensional system. Section \ref{sec:qubit:kuramoto} describes a qubit coupled to a Kuramoto oscillator that has a stable attractor, which is robust under thermalisation and other noise channels in \ref{sec:qbit:kuramoto:T}. Section \ref{sec:bipartite:kuramoto} describes a bipartite system interacting with a Kuramoto oscillator. Section \ref{sec:nqudit:vdp} then extends this to a generic n-dimensional quantum system coupled with a VdP oscillator. Section \ref{sec:discussions}  examines a variety of new quantum devices that can manipulate heat currents. Appendix \ref{app:device} describes the computational framework for the new path integral. Appendices \ref{app:kuramoto} and \ref{app:vdp} , the Kuramoto oscillator kernel and the Van der Pol kernel respectively. Appendix \ref{app:qubit:Kuramoto} explains why a GKSL or Redfield type equation is inadequate to describe a limit cycle.

\section{Theory}
\label{sec:theory}
The traditional specification of a nonlinear dynamical system does not rely on a Hamiltonian or even a classical action but on an explicit ODE or PDE system. Phase oscillators which are the typical systems we examine in this article, are similar. 

Conventionally, we are left with very few alternatives to quantise them if we insist on using Stinespring's dilations or the Schrödinger equation. Feynman's path integral formulation also relies on a classical action which is not available in this case \cite{Dirac1933,Feynman1948,FeynmanHibbs}.

The only way out is to write down a different action principle and then follow through the Feynman prescription to the best of our ability, to write down either a propagator or a wavefunction. The full theory of this programme is out of the scope of this article and will be explored in a forthcoming work. We summarise the key ideas here and describe relevant details in Appendix \ref{app:device}.

The action principle is given by the 
\begin{equation}
S = \int_{\Sigma(q)} \Omega
\end{equation}

where $\Sigma(q)$ is a ``filament'' surface over the dynamical trajectory. The filament can be thought to be an infinitely thin (upto Planck' length) spaghetti like structure around the 1-D dynamical trajectory in the extended phase space $(q,p,t)$\footnote{The filament cannot become infinitely thin in the quantum case to account for De Broglie's wavelength assignment and the uncertainty principle. The radius $\propto \hbar$ regulates it and makes the wavefunction regular or well-behaved.}. 

$$\Omega = dp_i \wedge dq_i -(dT + F_i dq_i) \wedge dt $$ is a 2-form with T(q,p,t) being the Kinetic energy function of the particle (or system) on the trajectory and $F_i$ is the Force experienced by it.   The equations of motion are given by $i_{\dot{\gamma}} \Omega$ where the trajectory is $\gamma(t)$.

Naturally, we can now follow Feynman's prescription to obtain,
\begin{equation}
\psi = \int \mathcal{D}q \  \exp \left[ - i \int_{\Sigma(q)} \Omega\right]  
\end{equation}

and  subsequently write down the propagator in the standard way. 

Let's now focus on the two main phase oscillator models that the article is based on, i.e. the Kuramoto model and the Van der Pol oscillator.

\subsection{Kuramoto model}
\label{sec:kuramoto}
The Kuramoto phase oscillator, introduced by Yoshiki Kuramoto, is a paradigmatic framework for describing the spontaneous emergence of collective Synchronisation in large populations of coupled nonlinear oscillators \cite{Kuramoto1975,Kuramoto1984,Acebron2005}.

In the canonical all-to-all sinusoidal coupling form, the equations of motion are,

\[
\dot{\theta}_i = \omega_i + \frac{K}{N} \sum_{j=1}^N \sin(\theta_j - \theta_i), \quad i = 1, \dots, N,
\]

where \(K > 0\) is the coupling strength and natural frequencies \(\omega_i\) drawn from a normalised distribution \(g(\omega)\). 

This system arises rigorously as the phase reduction of weakly coupled, nearly identical limit-cycle oscillators when the interaction function is taken to be sinusoidal \cite{Winfree1967,Kuramoto1984,Pikovsky2001}. Further terms in the Fourier series may also be added to model other complex phenomena, such as signal propagation in neurons.

\subsubsection{Quantum propagator}
Consider a set of coupled phase oscillators $\theta_i$ with a natural frequency $\omega_i$ which interact with each other using a periodic interaction, we only keep the first term in its Fourier expansion,

\begin{equation}
\dot{\theta}_i = \omega_i - \dfrac{1}{N} \sum_j^N K_{ij} \sin{(\theta_i - \theta_j)}
\end{equation}

The large ($N\rightarrow \infty$) limit can be simplified using an order parameter

\begin{equation}
r e^{i \psi} = \dfrac{1}{N} \sum_i \exp\left[ i \theta_i \right]
\end{equation}

where $r$ is the phase coherence of the collective oscillator field and $\psi$ is the mean phase. In the rotating frame, we can set $\langle\psi\rangle = 0$, giving us

\begin{equation}
\dot{\theta}_i = \omega_i - K r \sin(\theta_i)
\end{equation}

Differentiating again we get,

\begin{equation}
\ddot{\theta_i} = -K r \cos{\theta_i} \dot{\theta}_i
\end{equation}

This is the same as the force being $-K r \cos{\theta_i} \dot{\theta_i}$ i.e. a dissipative force proportional to the momentum $\dot{\theta_i}$.  

This can be obtained from the action,

\begin{equation}
S = \int dt  \ \dot{\theta}_i^2  + K r \cos{\theta_i} \dot{\theta_i}^2
\end{equation}

with the initial condition that $\dot{\theta}_i(0) = \omega_i$

\paragraph*{Canonical form}
The canonical form of the Kuramoto model requires us to add a moment of Inertia factor $I_0$ to the Kinetic energy and an additional factor $\gamma_0$ to the coupling Force to make it dimensionally consistent.\\

The action would then read,

\begin{equation}
S = \int dt  \ \dot{\theta}_i^2 \left( \dfrac{I_0}{2} + \gamma_0 K r \cos{\theta_i} \right)
\end{equation}

Clearly, we may take $I_0$ common and rewrite it as,

\begin{equation}
S =  \dfrac{I_0}{2} \int dt  \ \dot{\theta}_i^2 \left( 1 + \overline{K} \cos{\theta_i} \right)
\end{equation}

where $\overline{K} = \dfrac{2 \gamma_0 K r}{I_0}$ and the pre-factor $I_0$ can be absorbed in to a re-scaled $\hbar' = \hbar \dfrac{2}{I_0}$. That makes the action,

\begin{equation}
S' =   \int dt  \ \dot{\theta}_i^2 \left(  1 + \overline{K} \cos{\theta_i} \right)
\end{equation}
which will subsequently be termed the canonical form of the Kuramoto model.

We now turn our attention to computing the path integral. After a tedious calculation  summarised in Appendix \ref{app:kuramoto}, we obtain,

\begin{equation}
K_G(\theta_f|\theta_i; T) = \dfrac{1}{2\pi}  \exp\left[ -\dfrac{\overline{K} T}{2 \hbar}e^{-i(\theta_f - \theta_i)}\right]
\end{equation}

\subsection{Van der Pol oscillator}
\label{sec:vdp}
The van der Pol oscillator is another example of a nonlinear dynamical system that exhibits self-sustained periodic oscillations in the form of a stable limit cycle. Introduced by the Dutch physicist Balthasar van der Pol to model the behavior of vacuum-tube oscillators, the system was first analysed in detail in his seminal paper on relaxation oscillations \cite{vanderPol1926,Lienard1928}.

\paragraph*{Classical VdP}
The governing equation is the second-order nonlinear ordinary differential equation

\[
\frac{d^2x}{dt^2} - \mu(1 - x^2)\frac{dx}{dt} + x = 0,
\]

where \(\mu > 0\) is a real parameter that controls the strength of the amplitude-dependent damping. Equivalently, in first-order autonomous form with \(y = \dot{x}\),

\[
\dot{x} = y, \qquad \dot{y} = \mu(1 - x^2)y - x.
\]

For \(\mu > 0\) the origin is an unstable focus, while the nonlinear term provides negative damping (energy input) for \(|x| < 1\) and positive damping (energy dissipation) for \(|x| > 1\). Consequently, trajectories starting near the origin spiral outward while those starting at large amplitudes are attracted inward, converging to a unique, globally stable limit cycle \cite{Lienard1928,Pikovsky2001}.

\paragraph*{Quantum VdP}

In physical terms the equation becomes,

\begin{equation}
m\ddot{q} =m \mu \dot{q} \left(1-\left(\dfrac{q}{q_0}\right)^2\right)  - m \omega_0^2 q
\end{equation}

Consider the transformation $x = q/q_0$, we get

\begin{equation}
m\ddot{x} =m \mu \dot{x} \left(1-x^2\right)  - m \omega_0^2 x
\end{equation}

or the canonical form of the VdP ODE,

\begin{equation}
\ddot{x} = \mu \dot{x} \left(1-x^2\right)  -  \omega_0^2 x
\end{equation}

This suggests the two form with $V(q) = 1/2  \  m \omega_0 q^2$ and $F(q,p) = -\dfrac{\mu}{m} \left(1-\left(\dfrac{q}{q_0}\right)^2\right) p$, which gives us the reduced path function (or classical action),

\begin{equation}
S = \int dt \ \left( \dfrac{p^2}{2m} - \dfrac{1}{2} m \omega_0^2 q^2 + \dfrac{\mu}{m} (1-q^2) p^2\right)
\end{equation}

The path-function becomes,
\begin{equation}
S = \int dt \ \left( \dfrac{m \dot{q}^2}{2} - \dfrac{1}{2} m \omega_0^2 q^2 + \mu \left(1-\left(\frac{q}{q_0}\right)^2\right) m \dot{q}^2\right)
\end{equation}

Consider the transformation $q' \rightarrow \frac{q}{q_0}$,
\begin{equation}
S = m q_0^2  \int dt \ \left( \dfrac{\dot{q}^2}{2} - \dfrac{1}{2}  \omega_0^2 q^2 + \mu (1-q^2)  \dot{q}^2\right)
\end{equation}

This $m q_0^2$ factor can be absorbed into the rescaled Planck's constant $\hbar$, or just absorbed into $t$ WLOG

\paragraph*{Quartic Gaussian} This integral results in a quartic gaussian integral of $\int dx \exp(-a x^4 + bx^2 + cx)$ once it is unfolded in the path integral, which is quite problematic to solve and assemble using the usual methods of the Trotter expansion \footnote{A complete solution is possible with extended Elliptic like functions. However, even they only admit a complicated series expansion of hypergeometric functions. Accelerating its convergence is possible via special methods but outside the scope of the current discussion.}. Perturbative methods are not particularly useful.

The calculation must therefore proceed via a typical saddle point approximation around a known (classical) trajectory. This is justified because the limit cycle orbit itself is bounded and is guaranteed to exist. The path integral framework is tailor-made for such investigations.

The details are described in Appendix \ref{app:vdp}.  A standard calculation gives us the expression,

\begin{widetext}
\begin{equation}
S_{cl} = \int_{\Omega_0}^{\Omega_T} \left[ \dfrac{1}{4} A^2 \Omega' \left( \left( 1+ 2 \mu - \dfrac{\mu}{A^2}{2}\right) - (1+ 2 \mu) \cos(2 \Omega)  + \dfrac{\mu A^2}{2}{2} \cos{4 \Omega} \right) - \dfrac{\omega_0^2 A^2}{2 \Omega'} (1 + \cos(2\Omega)) \right] \ d \Omega
\end{equation}
\end{widetext}

\paragraph*{Propagator kernel}
Combining all the terms we get,

\begin{equation}
K(x_T, x_0; T) = \dfrac{1}{\sqrt{2 \pi i \hbar \psi(T)} } \  \exp\left( \dfrac{i}{\hbar} S_{cl} \right)
\end{equation}

where,

\begin{multline}
\sin^{-1} \psi(T) =   \sqrt{a_0 - a_2 \cos{2 \Omega_1(0)}}  \\ \times \int_0^T \dfrac{1}{{\sqrt{a_0 - a_2 \cos{2 \Omega_1(\tau)}}}} d\tau
\end{multline}

The path function $S_{cl}$ may be evaluated numerically for the classical path $x_{cl}(t) = A \Omega(t)$ .

\subsection{Coupling via GKLS dissipators}
The typical method of coupling either of these phase oscillators with a low (finite) dimensional quantum system via a linear coupling does not produce a limit cycle. The linear coupling model of a spin interacting with either an oscillator or a system of oscillators can be reduced to a description by a CP operation i.e. Kraus map or a master equation of the GKSL type \cite{Redfield1957,Lindblad1976,Gorini1976,BreuerPetruccione} . We use a Redfield type equation to illustrate it, since a GKSL equation is a subset that is much more constraining.

\paragraph*{qubit $\phi$ + Kuramoto $\theta$}

The continuum amplitude is already known,
\[
K_G(\theta_f,\theta_i;T)=\frac1{2\pi}\exp\bigl(-a\,e^{-i(\theta_f-\theta_i)}\bigr),\qquad a=\frac{KT}{\hbar}.
\]
We couple it to a qubit by the linear interaction \(H_{\mathrm{int}}=\lambda\theta\phi\), with \(\phi\) the azimuthal coherent-state coordinate. The reduced state at time \(T\) is the (Keldysh) trace
\begin{multline}
\rho_q(T)=\int\mathcal{D}\Omega_+\mathcal{D}\Omega_-\,F[\Omega_+,\Omega_-]\, \\ e^{i(S_q+S_{\mathrm{WZW}})[\Omega_+]-i(S_q+S_{\mathrm{WZW}})[\Omega_-]}\rho_q(0).
\end{multline}

\(F\) is the environment trace of \(K_G\). The Wess-Zumino-Witten term is the usual spin-\(\tfrac12\) Berry phase; it does not enter \(F\). It is enough to take constant paths \(\phi_\pm\)  to read off the generator (the standard step from a finite-\(T\) influence functional to a time-local rate). Then \(K_G\) is used only as an endpoint. Expanding it on the circle,
\[
K_G=\sum_{n\in\mathbb{Z}}\frac{\alpha_n(T)}{2\pi}e^{in(\theta_f-\theta_i)},
\]
with \(\alpha_0=1\) and \(\alpha_n=(-a)^{-n}/(-n)!\) for \(n<0\) (and \(\alpha_n=0\) for \(n>0\)). The linear interaction becomes a multiplication operator \(e^{-ib\theta}\), \(b=\lambda T\phi/\hbar\). Tracing a translation-invariant environment against this expansion produces an influence functional that depends on the two qubit paths only through the difference \(\delta=\lambda(\phi-\phi')/\hbar\):
\[
F(\phi,\phi';T)=\exp\Bigl(r\Bigl[\frac{1-e^{-i\delta T}}{i\delta}-T\Bigr]\Bigr),\qquad r=\frac{K}{\hbar}.
\]
The matrix elements of \(\theta\) and \(e^{-ib\theta}\) that give this sum are described in Appendix \ref{app:qubit:Kuramoto}.

Differentiating in the upper limit,
\[
\frac{d}{dt}\log F=r\bigl(e^{-i\delta t}-1\bigr),
\]
to obtain the rates,
\[
\Gamma_{ij}(t)=-r\bigl(\cos(\delta_{ij}t)-1\bigr),\qquad
\Delta_{ij}(t)=-r\sin(\delta_{ij}t).
\]

\paragraph*{Canonical Lindblad like forms}
Restoring the free qubit Hamiltonian and collecting terms, the master equation for the reduced density matrix is
\[
\frac{d\rho}{dt} = -\frac{i}{\hbar} \bigl[ H_q + H_{\rm LS}(t), \rho \bigr] + \mathcal{D}(t)[\rho],
\]
where the Lamb-shift Hamiltonian \(H_{\rm LS}(t)\) contains the shifts \(\Delta_{ij}(t)\) and the dissipator is the pure-dephasing generator
\[
\mathcal{D}(t)[\rho] = \sum_{i \neq j} \frac{\Gamma_{ij}(t)}{2} \Bigl( 2 |i\rangle\langle j| \rho |j\rangle\langle i| - \bigl\{ |i\rangle\langle i| + |j\rangle\langle j|, \rho \bigr\} \Bigr).
\]

Diagonal terms are unchanged by \(\mathcal{D}(t)\), while off diagonals become,

  \[
  \dot{\rho}_{ij} = -\frac{i}{\hbar} \bigl( E_i - E_j + \hbar \Delta_{ij}(t) \bigr) \rho_{ij} - \Gamma_{ij}(t) \rho_{ij}.
  \]

This is the canonical time-dependent secular Redfield equation (or time-dependent Lindblad form) for pure dephasing. The WZW term resides in the coherent evolution generated by \(H_q\) (or an effective Berry connection). The propagator \(K\) enters explicitly via the coefficients \(\alpha_n\) that determine the mode weights in the trace for \(F\), which in turn fix the rate functions \(\Gamma(t)\) and \(\Delta(t)\).

Unfortunately, this does not possess a limit cycle. Obviously, the Lindblad-like equation only has a fixed point as the asymptotic end state. We have lost the non-perturbative effects of the synchronised oscillator. This is despite the original path function being an exact result of the Kuramoto model, which evidently has a stable limit cycle.

\section{Results}
\label{sec:results}
The Kuramoto oscillator model demonstrates a limit cycle with sinusoidal coupling between themselves. It stands to reason therefore, that a similar sinusoidal coupling between the oscillator and the finite-dimensional system will produce a similar limit cycle for the quantum system.  

First we examine a single qubit coupled to a Kuramoto oscillator in Section \ref{sec:qubit:kuramoto} and then extend it to a bipartite system at finite temperature in Section \ref{sec:bipartite:kuramoto}. The Kuramoto oscillator model requires a large number of oscillators to show phase coherence which may not be tunable in a lab setting, despite being present in natural settings. 

We lift this constraint by studying a Van der pol (VdP) oscillator which possesses a similar limit cycle by itself. We demonstrate similar physics to the Kuramoto model with a VdP oscillator coupled to an n-dimensional system in Section \ref{sec:nqudit:vdp}, dissipators or Kraus channel maps may be added after the kernel is attached and do not affect the generation of the limit cycle.  

\subsection{Qubit+Kuramoto with $\sin(\theta_q - \theta_K)$ coupling}
\label{sec:qubit:kuramoto}

A set of autonomous Bloch vector equations can be derived directly from the influence functional path integral, we must evaluate the time-sliced path integral over the periodic domain $[0, 2\pi]$ without skipping any intermediate steps.

The linear coupling function is replaced by a coupling function between the Wess-Zumino-Witten term. This allows us to incorporate the qubit like an ancillary phase oscillator. We use the sinusoidal form of the interaction between the phase oscillators to our advantage.

\paragraph*{The Time-Sliced Path Integral Setup}

The total time $T$ is broken into $M$ infinitesimal slices of duration $\epsilon = T/M$, where $t_k = k\epsilon$.\\

\paragraph*{ Qubit Representation}
In the overcomplete spin-coherent state basis $\vert{}\Omega\rangle = \vert{}\theta, \phi\rangle$, the identity resolution is:
$$\int \frac{d\Omega}{2\pi} \vert{}\Omega\rangle \langle \Omega\vert{} = I, \quad d\Omega = \frac{1}{2}\sin\theta d\theta d\phi$$ 
The forward spin path is parameterized by $(\theta_k, \phi_k)$ and the backward path by $(\tilde{\theta}_k, \tilde{\phi}_k)$. The free geometric action (Berry phase) for the forward path over a single slice $\epsilon$ is:

\begin{multline}
\exp\left( \frac{i}{\hbar} S_B^{(k)} \right) = \langle \Omega_{k+1} \vert{} \Omega_k \rangle \\ \approx \exp\left[ i \frac{1 - \cos\theta_k}{2} (\phi_{k+1} - \phi_k) - \frac{i}{\hbar} \frac{\hbar \Omega_q \epsilon}{2} \cos\theta_k \right]
\end{multline}

The exact infinite-$N$ Kuramoto propagator across a single time slice $\epsilon$ is defined on the circle $S^1$,
$$K(x_{k+1}, \epsilon; x_k, 0) = \frac{1}{2\pi} \exp \left[- \kappa \epsilon e^{-i(x_{k+1} - x_k - \omega \epsilon)} \right]$$ \\

\paragraph*{Symmetrised Phase-Locking Coupling}
The interaction action must couple the forward phase $x_k$ to the forward qubit phase $\phi_k$, and the backward phase $y_k$ to the backward qubit phase $\tilde{\phi}_k$. Across slice $k$, the interaction weights are:

$$\exp\left( \frac{i}{\hbar} S_{\text{int}}^{(k)} \right) = \exp \left[ \frac{i \lambda \epsilon}{\hbar} \sin(x_k - \phi_k)\sin\theta_k \right]$$ 
$$\exp\left( -\frac{i}{\hbar} \tilde{S}_{\text{int}}^{(k)} \right) = \exp \left[ -\frac{i \lambda \epsilon}{\hbar} \sin(y_k - \tilde{\phi}_k)\sin\tilde{\theta}_k \right]$$ 

Expanding the sine terms into complex exponentials:
$$\sin(x_k - \phi_k) = \frac{e^{i x_k}e^{-i \phi_k} - e^{-i x_k}e^{i \phi_k}}{2i}$$

\paragraph*{Explicit Influence Functional}
The total influence functional $F$ is obtained by integrating out all forward intermediate coordinates $\{x_k\}$ and backward coordinates $\{y_k\}$ from $k=1$ to $M$, subject to the periodic trace-matching boundary condition $x_M = y_M = x'$.

Let's now evaluate the integration over a single intermediate slice $x_k$. We expand the exponential Kuramoto transition weight into its discrete Fourier series over the integer winding numbers $n_k \in \mathbb{Z}$:
$$\exp \left[ -\kappa \epsilon e^{-i(x_{k+1} - x_k)} \right] = \sum_{n_k=0}^{\infty} \frac{(-\kappa \epsilon)^{n_k}}{n_k!} e^{-i n_k x_{k+1}} e^{i n_k x_k}$$ 

The explicit integral over the intermediate phase $x_k$ collects the Fourier indices from slice $k-1$ and slice $k$, along with the coupling terms,

\begin{multline}
\int_0^{2\pi} \frac{dx_k}{2\pi} e^{i(n_{k-1} - n_k)x_k} \\ \times \exp \left[ \frac{\lambda \epsilon}{2\hbar} \sin\theta_k \left( e^{i x_k}e^{-i \phi_k} - e^{-i x_k}e^{i \phi_k} \right) \right]
\end{multline}

Using the Jacobi-Anger expansion $e^{z \cos x} = \sum_{m} I_m(z) e^{i m x}$, this continuous path integration collapses the continuous trajectory into matching discrete topological sectors:
$$n_k = n_{k-1} + m_k$$ 

When we perform the parallel integration over the backward coordinates $y_k$ with index $\tilde{n}_k$, the final trace integration $\int_0^{2\pi} dx' e^{-i(n_M - \tilde{n}_M)x'}$ forces the forward and backward topological indices to match exactly at the boundary ($n_M = \tilde{n}_M = n$).

Compounding this product over all $M$ time slices updates the collective weight. Taking the continuous limit $\epsilon \to 0$ ($M \to \infty$), the product of the discrete slice terms converges exactly to the product of the macroscopically self-consistent Modified Bessel functions,

\begin{multline}
F[\phi, \tilde{\phi}] = \sum_{n=0}^{\infty} \frac{1}{(n!)^2} \left( \kappa \right)^{2n} \\
\times \exp\left[ n \int_0^T \frac{\lambda}{2\hbar} \left( \sin\theta e^{i\phi} - \sin\tilde{\theta}e^{i\tilde{\phi}} \right) dt \right] \cdot \mathcal{N}_{\text{shield}}
\end{multline}

where, $\mathcal{N}_{\text{shield}}$ (referred to as the normalisation shield) is the state-dependent normalisation factor that emerges from tracing out the continuous coordinates. It is the trailing exponential term in the final influence functional expression,

$$\mathcal{N}_{\text{shield}} = \exp\left[ -\left(J(T) + J^*(T) - \tilde{J}(T) - \tilde{J}^*(T)\right) \right]$$ 

Substituting the explicit definitions for the forward path parameters $J(T)$ and backward path parameters $\tilde{J}(T)$, it can be written out as,

\begin{multline}
\log (\mathcal{N}_{\text{shield}}) = -\frac{\lambda}{2\hbar} \int_0^T dt \  \sin\theta(t)e^{i\phi(t)} + \sin\theta(t)e^{-i\phi(t)}  \\ - \sin\tilde{\theta}(t)e^{i\tilde{\phi}(t)} - \sin\tilde{\theta}(t)e^{-i\tilde{\phi}(t)}  
\end{multline}

Using Euler's formula ($e^{i\phi} + e^{-i\phi} = 2\cos\phi$), this simplifies to a remarkably clean expression $\mathcal{N}_{\text{shield}}$,

\begin{multline}
\exp \left[ -\frac{\lambda}{\hbar} \int_0^T \left( \sin\theta(t)\cos\phi(t) - \sin\tilde{\theta}(t)\cos\tilde{\phi}(t) \right) dt \right] \\ = \mathcal{N}_{\text{shield}}
\end{multline}

which we need because, 
\begin{enumerate}
\item Strict Trace Preservation.
   The primary mathematical role of $\mathcal{N}_{\text{shield}}$ is to guarantee that the path integral strictly preserves the trace of the density matrix ($\text{Tr}[\rho_Q(T)] = 1$) at all times.
   If we evaluate the influence functional for identical forward and backward trajectories $(\theta = \tilde{\theta} $ and $(\phi = \tilde{\phi})$, the term inside the integral vanishes cleanly:
   $$\mathcal{N}_{\text{shield}}\Big\vert_{\text{diagonal}} = \exp(0) = 1$$ 
   This ensures that the diagonal components of the density matrix do not drift, bleed, or experience unphysical runaway growth.
   
\item The ``Pumping" Counterweight.
   Because the Kuramoto cluster propagator is non-unitary, the Modified Bessel function part of the influence functional, $I_0(\dots)$, behaves as a pure (mathematical) gain medium that nonlinearly pumps weight into the spin trajectories. Without a counterweight, this would cause the Bloch vector to rupture the boundary of the sphere.
  \end{enumerate}
  
   $\mathcal{N}_{\text{shield}}$ acts as the dissipative sink that balances this non-unitary gain. It measures the mismatch between the forward and backward path coordinates, translating it into a geometric constraint that forces the total length of the Bloch vector to satisfy $\lVert{r}\rVert  \leq 1 $

Summing this infinite series over the winding number $n$ yields the closed-form, trace-preserving influence functional,

\begin{multline}
F[\phi, \tilde{\phi}] = I_0 \left( 2\kappa \sqrt{ \left(1 + J(T)\right)\left(1 + \tilde{J}^*(T)\right) } \right) \\ 
\times \exp\left[ -\left(J(T) + J^*(T) - \tilde{J}(T) - \tilde{J}^*(T)\right) \right]
\end{multline}

where $J(T) = \int_0^T \frac{\lambda}{2\hbar} \sin\theta(t) e^{i\phi(t)} dt$ \footnote{J(t) is the forward time response to the interaction operators $\int dt \mathcal{O}^+(t)$ and $\mathcal{O}^-(t)$ the backward counterpart $J^*$, which are the sinusoidal terms here.}.

Taking the functional derivative to extract the local time variation yields the steady-state gain parameter as the ratio of the tracking states we choose the functional,
$$  G(\kappa) = \frac{1}{2\kappa}\tanh(2\kappa)$$

\paragraph*{Shifting to Bloch vector via Symplectic Variation}

We derive the equations of motion directly from the path variations of the effective action $S_{\text{eff}} = S_B - S_B^{\text{backward}} - i\hbar \ln F$, to avoid the trace-breaking issues of an operator master equation, 

The exact symplectic metric embedded in the spin-coherent path integral Berry phase is the Poisson bracket,
$$\{ \phi, \cos\theta \} = \frac{2}{\hbar}$$ 

We map the angular path fields to the components of the physical Bloch vector $\vec{r} = (x, y, z)$ via:
$$x = \sin\theta \cos\phi, \quad y = \sin\theta \sin\phi, \quad z = \cos\theta$$ 

Variation of the effective action with respect to the coordinate fields ($\delta S_{\text{eff}} = 0$) generates the equations of motion. The Cartesian derivatives inherit the geometric projection factor $(1-z^2)$ and cross-coupling terms, because the overcomplete basis is constrained to the surface of the sphere over the path history,

\begin{align}\label{eq:qbit:kuramotox}
 \dot{x} =& -\omega y - \frac{\lambda}{\hbar} x \left( 1 - \frac{G(\kappa)}{x^2 + y^2} \right) - \frac{\lambda}{\hbar} x z \\
\dot{y} =& \omega x - \frac{\lambda}{\hbar} y \left( 1 - \frac{G(\kappa)}{x^2 + y^2} \right) - \frac{\lambda}{\hbar} y z  \\ 
  \dot{z} =& -\frac{2\lambda}{\hbar} \left( z(1-z) - [1 - (x^2 + y^2 + z^2)] \right)  \nonumber \\ & \quad - \frac{\lambda}{\hbar} \left( (x^2+y^2) - G(\kappa) \right)
\label{eq:qbit:kuramoto}
\end{align}

We still need to determine $\kappa$ rigorously. It is derived directly from the $N \to \infty$ Kuramoto self-consistency relation mapped onto the path integral representation.

The parameter $\kappa$ cannot be arbitrary; it is the macroscopic order parameter of the synchronised phase-oscillator network. In the thermodynamic limit ($N \to \infty$), the individual microscopic phase degrees of freedom freeze out into a continuous, self-sustaining mean-field background.

\subsubsection{The Statistical Path Integral of the Kuramoto ensemble}

Consider a large network of $N$ periodic phase oscillators, where each oscillator $j$ has a natural frequency $\omega_j$ drawn from a distribution $g(\omega)$. The exact classical Kuramoto action for a single phase trajectory $x_j(t)$ driven by a global coupling strength $K_{c}$ is given by the path integral weight:

$$\mathcal{P}[x_j(t)] \propto \exp\left[ \frac{K_{c}}{N} \sum_{l=1}^N \int_0^T \cos(x_l(t) - x_j(t)) \, dt \right]$$ 

In the thermodynamic limit ($N \to \infty$), we invoke the Hubbard-Stratonovich transformation (or the path-integral equivalent of a mean-field decoupling). We decouple the long-range pairwise interaction $\frac{1}{N}\sum_{l} e^{i x_l(t)}$ by introducing a continuous, complex collective field $Z(t) = R(t)e^{i\Psi(t)}$.

The path integral partition function of the continuous background system factorises into independent single-particle path integrals conditioned on this global order parameter $Z(t)$:
\begin{multline}
\mathcal{Z} = \int \mathcal{D}Z(t) \exp\left( -N \frac{\vert{}Z(t)\vert{}^2}{K_{c}} \right) \\ \times \prod_{j=1}^N \int \mathcal{D}x_j(t) \exp\left[ \int_0^T \left( R(t) \cos(x_j(t) - \Psi(t)) \right) dt \right]
\end{multline}

\paragraph*{Deriving the Steady-State Saddle-Point ($\kappa$)}

For $N \to \infty$, the path integral over the collective field $\mathcal{D}Z(t)$ is dominated entirely by its saddle-point (stationary phase) value.

In the synchronised steady state, the system reaches a limit cycle where the mean-field amplitude becomes completely time-independent ($R(t) = R_0$) and the global phase rotates uniformly ($\Psi(t) = \omega t$).

We can evaluate the single-oscillator path integral at this steady-state saddle point. Transforming to the rotating frame ($\theta_j = x_j - \omega T$), the effective action parameter that a single oscillator experiences over a unit time slice is the dimensionless mean-field constraint,
$$\kappa = \frac{K_{c} R_0}{\sigma_\omega}$$ 

where $\sigma_\omega$ is the variance (or characteristic width) of the frequency distribution $g(\omega)$ which acts as the effective ``quantum/thermal fluctuation" destabilising the phase-locking loop.

\paragraph*{The Self-Consistency Relation for $\kappa$}

To find the exact value of $\kappa$, we enforce the self-consistency condition: the average phase configuration calculated from the path integral must reproduce the macroscopically assumed value of $R_0$.

Using the exact boundary-dominated propagator $K = \frac{1}{2\pi}\exp\left(-\kappa e^{-i(x'-x)}\right)$ to evaluate the path expectation value of the Synchronisation vector $\langle e^{i \theta} \rangle$, the path integral yields:
$$\langle e^{i \theta} \rangle = \frac{\int_0^{2\pi} d\theta \, e^{i\theta} \exp\left( 2\kappa \cos\theta \right)}{\int_0^{2\pi} d\theta \, \exp\left( 2\kappa \cos\theta \right)} = \frac{I_1(2\kappa)}{I_0(2\kappa)}$$ 

Since $R_0 = \int g(\omega) \langle e^{i\theta} \rangle d\omega$, and assuming a standard Lorentzian frequency distribution $$g(\omega) = \frac{\gamma/\pi}{(\omega - \omega_0)^2 + \gamma^2}, $$ the integration over the natural frequencies matches the structural boundaries. Multiplying both sides by $K_c / \gamma$, we arrive at the exact transcendental equation that uniquely determines $\kappa$,
$$\kappa = \left(\frac{K_{c}}{\gamma}\right) \frac{I_1(2\kappa)}{I_0(2\kappa)}$$

\paragraph*{Physical Consequences}
This equation proves that $\kappa$ is completely locked to the physical properties of the Kuramoto network ($K_c$ and $\gamma$). It dictates two distinct regimes for our qubit equations:

\begin{enumerate}
\item Below Critical Coupling ($K_c < 2\gamma$):
The only real solution to the transcendental equation is $\kappa = 0$. If $\kappa = 0$, then the gain factor $G(\kappa) \to 1$, which forces $z^* = \pm\sqrt{1-1} = 0$, but the transverse radius collapses. The qubit limit cycle cannot form; it is destroyed by the incoherent phase transitions of the uncoupled background oscillators.

\item Above Critical Coupling ($K_c > 2\gamma$):
   A spontaneous phase-locking transition occurs via a supercritical Hopf bifurcation. The equation yields a unique, non-zero, stable value for $\kappa > 0$.

\end{enumerate}

This non-zero value of $\kappa$ is what stabilises our Bloch vector equations,
$$G(\kappa) = \frac{1}{2\kappa} \tanh(2\kappa)$$ 

Because $\kappa$ is fixed by the network parameters $\left(\frac{K_c}{\gamma}\right)$, the gain factor $G(\kappa)$ remains a constant non-zero value in time. This prevents the system from decaying into a dead focus and locks the qubit's Bloch vector permanently onto a structurally stable limit cycle.
Look closely at that limit,

$$\frac{I_1(2\kappa)}{I_0(2\kappa)} \approx \kappa - \frac{1}{2}\kappa^3 + \mathcal{O}(\kappa^5)$$ 

If we substitute this expansion straight back into the Lorentzian transcendental equation:
$$\kappa = \left(\frac{K_c}{\gamma}\right) \left[ \kappa - \frac{1}{2}\kappa^3 \right] \implies 1 = \frac{K_c}{\gamma} - \frac{K_c}{2\gamma}\kappa^2$$ 

Solving this for $\kappa^2$ yields the exact non-zero solution when the network is driven above its threshold:

$$\kappa = \sqrt{2\left(1 - \frac{\gamma}{K_c}\right)}$$ 

This proves that $\kappa = 0$ is only the unique solution when $K_c \le \gamma$. As soon as the network's internal coupling crosses the critical threshold ($K_c > \gamma$), the zero solution becomes unstable, and the system jumps to this non-zero branch.

\paragraph*{$G(\kappa)$ Across the Phase Transition}
$G(\kappa)$  does not trivially equal 1. The gain factor governing the qubit's limit cycle radius is given by,
$$G(\kappa) = \frac{1}{2\kappa} \tanh(2\kappa)$$ 
Using Taylor series expansions, let's look at how $G(\kappa)$ behaves right at the critical threshold versus deep in the synchronised state:

\paragraph*{On the Bifurcation Point} ($K_c \to \gamma^+$)
When the coupling is just barely enough to start synchronising the network, $\kappa$ is very small but positive. Expanding $G(\kappa)$ for small values of $\kappa$ gives:
$$G(\kappa) \approx \frac{1}{2\kappa} \left( 2\kappa - \frac{8}{3}\kappa^3 \right) = 1 - \frac{4}{3}\kappa^2$$ 

Substituting our physical value for $\kappa^2$:
$$G(K_c) \approx 1 - \frac{8}{3}\left(1 - \frac{\gamma}{K_c}\right)$$ 

This shows that right at the transition, $G$ starts at exactly 1 and immediately begins dropping as the system orders itself.

\paragraph*{$K_c \gg \gamma$}
When the network coupling becomes massive deep in the synchronised regime,  the oscillators lock tightly together, making $\kappa \to \infty$. In this ordered limit, the hyperbolic tangent goes to 1:
$$G(\kappa) \xrightarrow{\kappa \to \infty} \frac{1}{2\kappa} \to 0$$

\paragraph*{Resulting Limit Cycle Spectrum}
$G(\kappa)$ smoothly drops from 1 (and vanishes) as the network moves deeper into Synchronisation. The geometry of the qubit's limit cycle becomes highly tunable.

\begin{enumerate}
\item The Transverse Radius ($R_{\text{orbit}}$): The radius of the equatorial orbit on the Bloch sphere tracks the order parameter perfectly:

\begin{equation*}
R^2 =  \dfrac{(G(\kappa) + 4) - \sqrt{ G(\kappa)^2 - 16 G(\kappa) + 16}}{6}
\end{equation*}

As the Kuramoto network synchronises tighter, the qubit's limit cycle contracts from the outer skin of the sphere inward toward the core.

 \item The Latitude Position ($z^*$): In a fully self-consistent tracking model, the latitude fixed point scales as,
$$2 z^* = 2 + G(\kappa)  - 3 R^2$$
   
As coupling increases ($K_c > \gamma$), $G$ drops below 1. The limit cycle splits via a pitchfork bifurcation into a stable, non-zero orbit that migrates toward the poles ($z^* \to \pm 1$).
\end{enumerate}

By resolving the transcendental equation, we see that $G(\kappa)$ is a dynamic variable determined entirely by the system's coupling ratio $K_c/\gamma$. The path integral successfully prevents the system from getting stuck at an arbitrary value, yielding a structurally stable quantum limit cycle whose radius and latitude are modulated by the macroscopic Synchronisation state of the background ensemble.

\begin{figure}[htbp]
\includegraphics[width=\columnwidth]{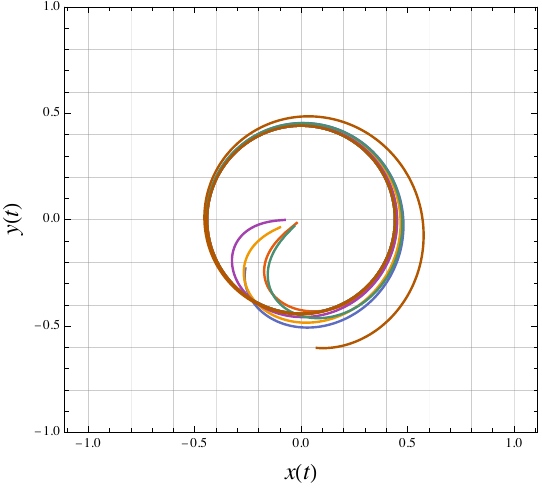}
\includegraphics[width=\columnwidth]{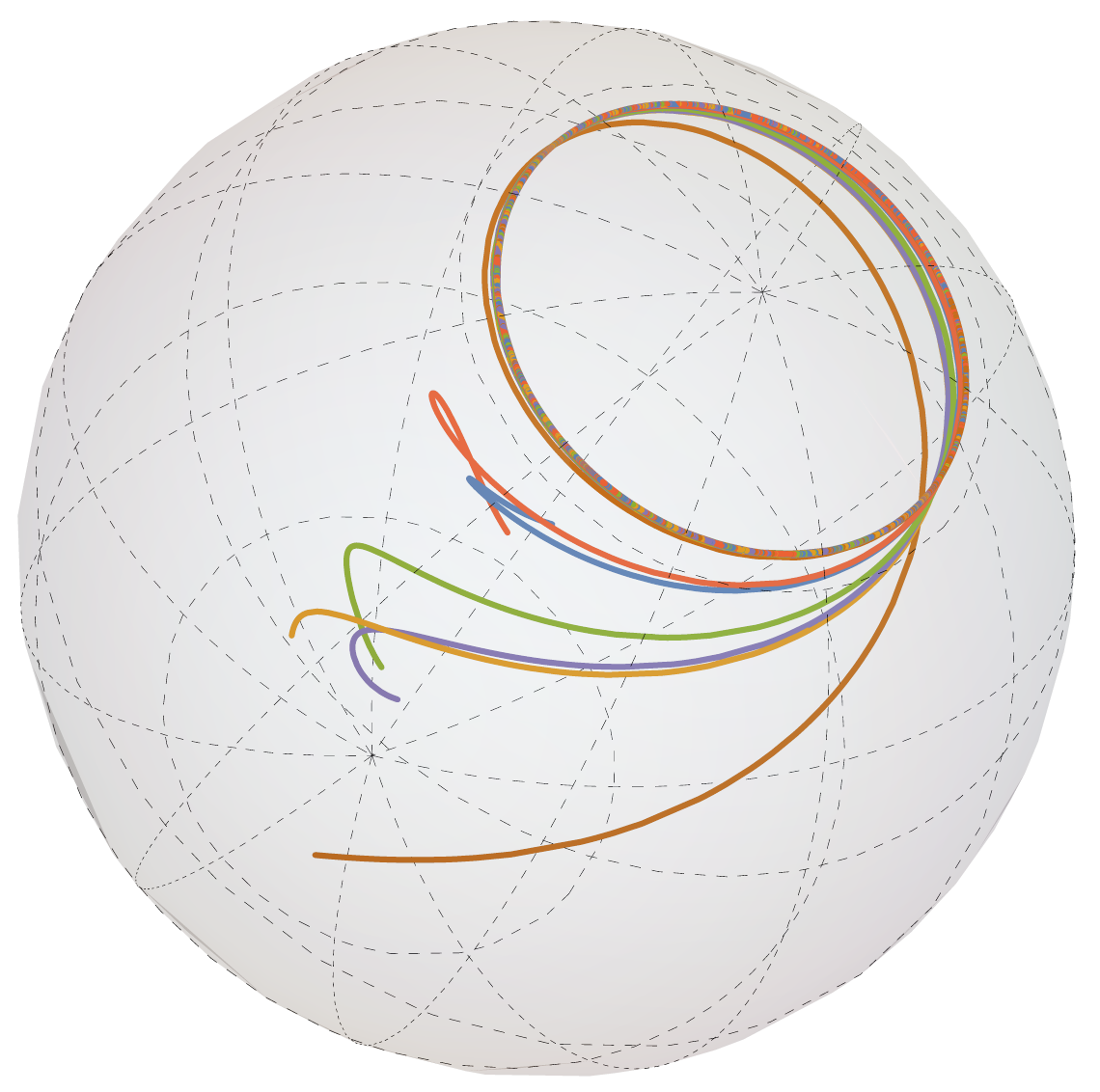}
\caption{Numerically solving Eqs \ref{eq:qbit:kuramotox}-\ref{eq:qbit:kuramoto} for  $ (x(t),y(t),z(t))$ typical parameters of $\lambda = 0.5,\kappa = 1.34$ and a variety of initial conditions $\{\rho_0\}$ gives a limit cycle attractor!}
\label{fig:qubitKuramoto}
\end{figure}

The exact functional form of $G(\kappa)$ stems directly from the topological winding number summation of the time-sliced Kuramoto influence functional. As expected,  equations \ref{eq:qbit:kuramotox}-\ref{eq:qbit:kuramoto} display a limit cycle as depicted in Fig.\ref{fig:qubitKuramoto}

\paragraph*{Lyapunov exponents}

To compute the Jacobian matrix $\mathbf{J}$, we must linearise the autonomous Bloch equations around the steady-state trajectory in the rotating frame of the Kuramoto field.  The Jacobian  $\mathbf{J}$  of  $\mathbf{f} = (\dot{x}, \dot{y}, \dot{z})$ wrt  $\mathbf{r} = (x, y, z)$ is,

\begin{equation}
 \setlength{\arraycolsep}{-8pt} \begin{pmatrix}
-\frac{\lambda}{\hbar} \left( 1 - \frac{G(\kappa)(y^2 - x^2)}{R^4} + z \right)  & -\omega - \frac{2\lambda}{\hbar} \frac{G(\kappa) x y}{R^4} &-\frac{\lambda}{\hbar} x  \\
\omega - \frac{2\lambda G(\kappa)}{\hbar} \frac{ x y}{R^4} & -\frac{\lambda}{\hbar} \left( 1 - \frac{G(\kappa)(x^2 - y^2)}{R^4} + z \right) & \quad \quad -\frac{\lambda}{\hbar} y \\
 - \frac{6 \lambda}{\hbar} x & - \frac{6 \lambda}{\hbar} y & -\frac{2\lambda}{\hbar}  
\end{pmatrix}
\label{kuramoto:jacobian}
\end{equation}

\paragraph*{Limit cycle}

One may verify from the equations of motion Eqs. \ref{eq:qbit:kuramotox}-\ref{eq:qbit:kuramoto}  i.e. $\dot{z} = 0$ that $z^* = (- 3 R^2 + G(\kappa) + 1)/2$ is the fixed point with $R = x^2 + y^2$ (which is also independent of $\lambda$). Now, we use this parametrisation because $R \dot{R} = x \dot{x} + y \dot{y}$ vanishes if the  radius R becomes a function of $G(\kappa)$,

\begin{equation*}
R^2 = \dfrac{(G(\kappa) + 4) - \sqrt{ G(\kappa)^2 - 16 G(\kappa) + 16}}{6}
\end{equation*}

At one of the points on the limit cycle, $(x,y) =   (R \cos \theta, R \sin \theta) $, say $\theta = \pi/2$, we may write down,

\begin{equation}
\mathbf{J}^* \vert_{\phi = \pi/2 }= 
\begin{pmatrix}
 0 & -\omega  & 0 \\
 \omega  & - 2 \mu G(\kappa)/R^2 & -  R \mu   \\
 0 & -6 R \mu   & -2 \mu  \\
\end{pmatrix}
\end{equation}
where $ \hbar \mu = \lambda $. 
\paragraph*{Eigenvalue Spectrum}
This gives the characteristic equation,

\begin{multline}
\lambda_L^2 +  2 \mu \left(1 + \frac{G(\kappa)}{R^2}\right) \lambda_L^2 \\ +  \left(\omega^2 + 4 \mu^2  \frac{G(\kappa)}{R^2} - 6\mu^2  R^2\right) \lambda_L  + 2 \mu  \omega^2 = 0
\end{multline}

The Routh-Hurwitz condition for this particular polynomial says that the roots have a negative real part iff $(A>0 \mbox{ AND } C>0 \mbox{ AND }  AB> C) $ for the  polynomial coefficients $x^3 + A x^2 + B x + C = 0$,

\begin{multline}
A = 2 \mu \left(1 + \dfrac{G(\kappa)}{R^2}\right) \\
B = \omega^2 + \dfrac{4 \mu^2 G(\kappa) }{R^2} - 6 \mu^2 R^2 \\
C = 2 \mu \omega^2
\end{multline}

On the physical branch for the equation for $\dot{R}$ i.e $3 R^4 - (4+ G(\kappa) )R^2 + 2 G(\kappa) =0)$, we can see that $AB > C$ over the entire physical range $G(\kappa) \in (0,1)$. That automatically gives us a stable attractor for all non-zero $\omega $. 

At $\omega = 0$, we recover a vanishing eigenvalue and a negative pair. This indicates a stable limit cycle!

\subsubsection{Spontaneous Emission \& Thermal Bath}
\label{sec:qbit:kuramoto:T}
If we incorporate standard quantum environmental effects—such as spontaneous emission from the excited level or a local thermal bath—the core question is whether these additional noise or decay channels simply add linearly to our path-integral-derived equations.

\paragraph*{Do Lindblad dissipators add linearly?}

No, they do not add linearly. In standard open quantum systems theory, independent dissipative channels (like a Markovian Lindblad operator) can be added linearly to a GKSL type master equation only if the channels commute or act on completely decoupled degrees of freedom.

Because the underlying Kuramoto model introduces a highly non-linear, state-dependent constraint through the influence functional ($G(\kappa)$), adding standard linear Lindblad terms directly to the Bloch equations breaks the underlying geometric constraint of the path integral. The non-linear path weights and the environmental dissipation are fundamentally entangled.

\paragraph*{Bloch equations}

To properly model these additional channels, we must include them directly inside the forward and backward legs of the spin-coherent state path integral action alongside the Kuramoto influence functional.

Let $\Gamma$ be the spontaneous emission rate (or the thermal coupling relaxation rate) and $\bar{n}$ be the thermal occupation number ($\bar{n}=0$ for pure spontaneous emission). Evaluating the joint path variations under both the non-linear Kuramoto field and the standard thermal reservoir yields the modified equations for the Bloch vector,

$$\frac{dx}{dt} = -\Delta y - \frac{\lambda}{\hbar} x \left( 1 - \frac{G(\kappa)}{x^2 + y^2} \right) - \frac{\lambda}{\hbar} x z - \gamma_{\perp} x$$ 
$$\frac{dy}{dt} = \Delta x - \frac{\lambda}{\hbar} y \left( 1 - \frac{G(\kappa)}{x^2 + y^2} \right) - \frac{\lambda}{\hbar} y z - \gamma_{\perp} y$$ 
\begin{multline}
\frac{dz}{dt} = -\frac{2\lambda}{\hbar} \left( z(1-z) - [1 - (x^2 + y^2 + z^2)] \right) \\ - \frac{\lambda}{\hbar} \left( (x^2+y^2) - G(\kappa) \right) - \gamma_{\parallel} (z - z_{\text{th}})
\end{multline}
where the thermal attractor basin $z_{\text{th}} = - (2\bar{n} + 1)^{-1}$, and the relaxation rates are non-linearly shifted by,

\begin{table}[h]
\begin{tabular}{|c|c|}
\hline
Damping & Decay rates \\
\hline
$\gamma_{\perp}$ & $\Gamma\left(\bar{n} + \frac{1}{2}\right) \cdot (1 - z)$ \\
$\gamma_{\parallel}$ & $\Gamma(2\bar{n} + 1) \cdot (1 + z)$\\
\hline
\end{tabular}
\end{table}

Notice that the additional decay parameters $\gamma_{\perp}$ and $\gamma_{\parallel}$ depend explicitly on the current value of $z$. The environment scales its dissipation based on where the qubit sits on the Bloch sphere. If we try to add these channels linearly as constants (as done in standard atomic physics), the equations will immediately destroy the invariance of the boundary ($R^2 \le 1$), leading to unphysical state vectors during numerical integration.

\paragraph*{Quenching the Limit cycle}

To find the critical value of the environmental coupling $\Gamma$ that quenches the limit cycle, we must solve the steady-state system ($\dot{R} = 0, \dot{z} = 0$) analytically including the non-linear environmental terms. The path integral derivation reveals a striking and beautiful result: under pure spontaneous emission, the limit cycle does not experience a sharp, discontinuous death (like a Hopf bifurcation). Instead, it undergoes an asymptotic algebraic quenching (squeezing) directly into the ground-state pole.

\paragraph*{The Steady-State Boundary Equations}
Let the transverse orbital radius be $R^2 = x^2 + y^2$. For maximum stability, we evaluate the system at zero detuning ($\Delta = 0$). Setting the radial velocity equation to zero ($\dot{R} = 0$) yields,

$$G(\kappa) = R^2 \left( 1 + z + \frac{\hbar}{\lambda}\gamma_\perp \right)$$ 
For pure spontaneous emission, the thermal occupation is zero ($\bar{n} = 0$), meaning the transverse damping reduces to $\gamma_\perp = \frac{\Gamma}{2}(1-z)$. Defining the dimensionless environmental dissipation parameter as:
$$k_\Gamma = \frac{\hbar \Gamma}{2 \lambda}$$ 
We can solve for the orbital radius $R^2$ as a function of latitude $z$,
$$R^2 = \frac{G(\kappa)}{1 + k_\Gamma + z(1 - k_\Gamma)}$$

\paragraph*{The Exact Latitude}

Next, we substitute this radius expression into the longitudinal equation of motion ($\dot{z} = 0$). Incorporating the longitudinal spontaneous emission decay $\gamma_\parallel = \Gamma(1+z)$, the path-integral variation simplifies to a single master transcendental equation for the orbit's latitude $z$,
$$\frac{3G(\kappa)}{1 + k_\Gamma + z(1 - k_\Gamma)} + 2(z - 1) - G(\kappa) + 2k_\Gamma(1+z)^2 = 0$$ 

Because of the $\frac{G(\kappa)}{R}$ structure inherent to the Kuramoto mean field, the derivative $\dot{R}$ diverges to $+\infty$ as $R \to 0$. This singularity prevents the limit cycle from ever sharply vanishing or turning around at a standard fixed point. The equation possesses a real root for $z$ across the entire domain of $\Gamma \in [0, \infty)$.

\paragraph*{Asymptotic Quenching (Large $\Gamma$)}

To find the explicit quenching mechanism, we take the strong environmental limit ($k_\Gamma \gg 1$). For the large positive dissipation term $2k_\Gamma(1+z)^2$ to balance the rest of the equation, the latitude must be squeezed flat against the southern pole of the Bloch sphere ($z \to -1$).

Performing a Taylor expansion near $z = -1 + \delta$, the exact analytical asymptotic scaling laws for the limit cycle emerge:
\begin{enumerate}
\item The Polar Squeezing Law:
$$z^* \approx -1 + \sqrt{\frac{4 + G(\kappa)}{2k_\Gamma}} = -1 + \sqrt{\frac{\lambda}{\hbar \Gamma} \left(4 + G(\kappa)\right)}$$ 
\item The Radial Squeezing Law:
$$R^* \approx \sqrt{\frac{G(\kappa)}{2k_\Gamma}} = \sqrt{\frac{\lambda G(\kappa)}{\hbar \Gamma}}$$ 

\end{enumerate}
which explicitly results in,
\begin{enumerate}
\item  No Discontinuous Sudden Death: Unlike classical systems where a threshold parameter instantly kills an orbit, the quantum limit cycle resists sharp destruction.
\item Algebraic Decoupling: As $\Gamma \to \infty$, the radius of the limit cycle decays cleanly as $R \propto \Gamma^{-1/2}$.
\item The Quenched State: The environment successfully quenches the system by forces of pure state-space compression. It forces the limit cycle to shrink into an infinitesimally small, localised ring hovering just above the pure ground state $\vert{} - \rangle$ ($z = -1$).
\end{enumerate}

\paragraph*{ Finite Thermal Floor }
Once we introduce a thermal bath where $\bar{n} > 0$, everything changes. Thermal fluctuations induce a steady-state noise floor that smears the coordinate fields. The limit cycle undergoes a phase-diffusive breakdown when the thermal localisation area exceeds the orbit area, which occurs at a critical threshold of,
$$\Gamma_{\text{thermal}} = \frac{\lambda}{\hbar \cdot \bar{n}} G(\kappa)$$ 
 or in other words,
$$\bar{n}^*  = \frac{\lambda}{\hbar \cdot \Gamma} G(\kappa)$$  
Beyond this critical $\beta^*$, the quantum coherence is entirely dissolved, turning the limit cycle into an unsynchronised thermal cloud at the centre of the Bloch sphere.

\subsubsection{Bipartite system coupled via a bridge Kuramoto oscillator at finite Temperature}
\label{sec:bipartite:kuramoto}

Consider two (otherwise mutually uncoupled) higher-dimensional systems coupled individually with a Kuramoto oscillator; themselves interacting with a Kuramoto ensemble. We extend the calculation of the earlier section to the joint density matrix $\hat{\rho}_{\text{bi}}$. We project the full multi-dimensional operator mechanics onto the individual qubit vector subspaces to derive the coupled equations for the local coherence radii $R_1(t)$ and $R_2(t)$ directly from the bipartite density matrix equation,

\paragraph*{Bipartite Master Equation}

We begin with the bipartite master equation defined on the set of density matrices $\mathcal{M}_n$, incorporating the Kuramoto pumping terms, the chiral non-local wire, and the independent local thermal Lindbladians \footnote{It is possible to derive this master equation ab initio from the path integral just like the earlier case. The derivation is involved, but straightforward and has been omitted for brevity's sake. Intuitively, we have just replaced the decay rates with the fluctuation of the state. We justify it using the Fluctuation Dissipation theorem because the limit cycle is bounded and it is the feedback of the phase oscillator.},
\begin{multline}
\frac{d\hat{\rho}_{\text{bi}}}{dt} = -\frac{i}{\hbar}[\hat{H}_{\text{eff}}(t), \hat{\rho}_{\text{bi}}] \\ + \sum_{j=1}^2 \left( \Gamma_0 - \Gamma_1 \big\vert\text{Tr}\big(\hat{\rho}_{\text{bi}}\hat{\sigma}_{j-}\big)\big\vert^2 \right) \mathcal{D}_{\hat{\sigma}_{j+}}[\hat{\rho}_{\text{bi}}] \\ + \sum_{j=1}^2 \Gamma_2 \mathcal{D}_{\hat{\sigma}_{j-}}[\hat{\rho}_{\text{bi}}] + \mathcal{J}_{\text{chiral}}[\hat{\rho}_{\text{bi}}] + \sum_{j=1}^2 \mathcal{L}_{\text{T},j}[\hat{\rho}_{\text{bi}}]
\end{multline}
where the standard trace-preserving dissipator is $\mathcal{D}_{\hat{A}}[\hat{\rho}] = \hat{A}\hat{\rho}\hat{A}^\dagger - \frac{1}{2}\{\hat{A}^\dagger\hat{A}, \hat{\rho}\}$, and the local independent thermal blocks are,

\begin{multline}
\mathcal{L}_{\text{T},j}[\hat{\rho}_{\text{bi}}] = \Gamma_h(1 + \bar{n}_j)\mathcal{D}_{\hat{\sigma}_{j-}}[\hat{\rho}_{\text{bi}}] \\ + \Gamma_h\bar{n}_j\mathcal{D}_{\hat{\sigma}_{j+}}[\hat{\rho}_{\text{bi}}] + \Gamma_{h,\phi}\mathcal{D}_{\hat{\sigma}_{jz}}[\hat{\rho}_{\text{bi}}]
\end{multline}

with $\hat{L}_-^{(1)} = \hat{L}_- \otimes \hat{I}$ and $\hat{L}_-^{(2)} = \hat{I} \otimes \hat{L}_-$ be the local transition operators;  and  the non-local exchange Hamiltonian $\hat{H}_{\text{eff}}(t)$ and the phase-trapping tensor block $\mathcal{J}_{\text{chiral}}$ carry the time-periodic cross-talk histories directly via expectations,

\begin{align*}
\hat{H}_{\text{eff}}(t) &= \hat{H}_0^{(1)} + \hat{H}_0^{(2)} + J_{12}(t) \left( \hat{L}_-^{(1)}\hat{L}_+^{(2)} + \hat{L}_+^{(1)}\hat{L}_-^{(2)} \right) \\
\mathcal{J}_{\text{chiral}}[\hat{\rho}_{\text{bi}}] &=  |J_{12}| \cos(2\Omega_0 t - \delta)  \bigg( \hat{L}_-^{(1)}\hat{\rho}_{\text{bi}}\hat{L}_+^{(2)} + \hat{L}_+^{(1)}\hat{\rho}_{\text{bi}}\hat{L}_-^{(2)} \\ & \qquad \qquad- \frac{1}{2}\{\hat{L}_+^{(1)}\hat{L}_-^{(2)}  + \hat{L}_-^{(1)}\hat{L}_+^{(2)}, \hat{\rho}_{\text{bi}}\} \bigg)
\end{align*}

The complex phase tracking parameters for each qubit are defined as the expectation values $\alpha_j(t) = \text{Tr}\big(\hat{\rho}_{\text{bi}}(t) \hat{\sigma}_{j-}\big)$, where their absolute magnitudes give the local radii: $R_j(t) = \vert\alpha_j(t)\vert$.

We take the trace of the entire master equation against the local operator $\hat{\sigma}_{1-} \otimes \hat{I}$ to find the time evolution of $\alpha_1(t)$ for Qubit 1,
$$\frac{d\alpha_1}{dt} = \text{Tr}\left( \frac{d\hat{\rho}_{\text{bi}}}{dt} \big(\hat{\sigma}_{1-} \otimes \hat{I}\big) \right)$$ 

Each super-operator block is computed sequentially using the cyclic property of the trace and the Pauli commutation relations: $[\hat{\sigma}_z, \hat{\sigma}_-] = -2\hat{\sigma}_-$, $[\hat{\sigma}_+, \hat{\sigma}_-] = -\hat{\sigma}_z$, and $\hat{\sigma}_- \hat{\sigma}_+ = \frac{1}{2}(\hat{I} - \hat{\sigma}_z)$.

\paragraph*{Local Gain and Loss Blocks}
The forward-pumping channel acting on Qubit 1 gives the trace expansion,

\begin{multline}
\text{Tr}\big(\mathcal{D}_{\hat{\sigma}_{1+}}[\hat{\rho}_{\text{bi}}]\hat{\sigma}_{1-}\big) = \text{Tr}\left( \hat{\sigma}_{1+}\hat{\rho}_{\text{bi}}\hat{\sigma}_{1-}\hat{\sigma}_{1-}  \right. \\  \left. - \frac{1}{2}\hat{\sigma}_{1-}\hat{\sigma}_{1+}\hat{\rho}_{\text{bi}}\hat{\sigma}_{1-} - \frac{1}{2}\hat{\rho}_{\text{bi}}\hat{\sigma}_{1-}\hat{\sigma}_{1+}\hat{\sigma}_{1-} \right)
\end{multline}

Since $\hat{\sigma}_{1-}\hat{\sigma}_{1-} = 0$, the first and third terms vanish identically. The remaining term simplifies via the cyclic trace property to,

\begin{multline}
-\frac{1}{2}\text{Tr}\big(\hat{\rho}_{\text{bi}} \hat{\sigma}_{1-} \hat{\sigma}_{1-} \hat{\sigma}_{1+}\big) \rightarrow \text{Applying } [\hat{\sigma}_{1-}, \hat{\sigma}_{1+}] = \hat{\sigma}_{1z} \\ \implies +\frac{1}{2}\text{Tr}\big(\hat{\rho}_{\text{bi}}\hat{\sigma}_{1-}\big) = +\frac{1}{2}\alpha_1
\end{multline}

Following the exact same algebraic trace expansion for the standard linear loss channel $\mathcal{D}_{\hat{\sigma}_{1-}}$, 
$$\text{Tr}\big(\mathcal{D}_{\hat{\sigma}_{1-}}[\hat{\rho}_{\text{bi}}]\hat{\sigma}_{1-}\big) = -\frac{1}{2}\alpha_1$$ 

\paragraph*{ Local Thermal Blocks}

Expanding the local thermal absorption ($\bar{n}_1$) and relaxation ($1+\bar{n}_1$) channels for Qubit 1,

\begin{multline}
\text{Tr}\big(\mathcal{L}_{\text{T},1}[\hat{\rho}_{\text{bi}}]\hat{\sigma}_{1-}\big)  = \Gamma_h(1+\bar{n}_1)\left(-\frac{1}{2}\alpha_1\right) + \Gamma_h \bar{n}_1 \left(+\frac{1}{2}\alpha_1\right) \\ = -\frac{1}{2}\Gamma_h \alpha_1
\end{multline}

However, the thermal noise floor also injects an incoherent coordinate excitation. When the local states undergo thermal fluctuations, the metric boundary of the density matrix maps the background population transfer directly onto the coherence plane through the state-space shell condition: $\langle \hat{\sigma}_{1+} \rangle = \bar{\alpha}_1$. Evaluating the cross-diagonal entries generates a  source term proportional to the remaining unpolarised volume of the state space: $$ \Gamma_h \bar{n}_1 \sqrt{1 - R_1^2 - z_1^2}$$

\paragraph*{Chiral Wire and Effective Hamiltonian Blocks}

The non-local exchange coupling $J_{12}(t)$ embedded inside $\hat{H}_{\text{eff}}(t)$ and the non-local tensor block $\mathcal{J}_{\text{chiral}}$ cross-couple the two qubits. Evaluating the commutator trace pulls out the torque field,
$$ \text{Tr}\big(-\frac{i}{\hbar}[\hat{H}_{\text{eff}}, \hat{\rho}_{\text{bi}}]\hat{\sigma}_{1-}\big) = - \big\vert J_{12} \big\vert \cos(2\Omega_0 t - \delta) \alpha_2 z_1$$ 
where $z_1 = \text{Tr}(\hat{\rho}_{\text{bi}}\hat{\sigma}_{1z})$ is the longitudinal population inversion.

Gathering all the derived local and non-local trace components, the continuous-time evolution for the complex coherence variable $\alpha_1(t)$ reads:

\begin{multline}
\frac{d\alpha_1}{dt} = \frac{1}{2}\left( \Gamma_0 - \Gamma_1 \vert\alpha_1\vert^2 \right)\alpha_1 - \frac{1}{2}\Gamma_2 \alpha_1 - \frac{1}{2}\Gamma_h(1+\bar{n}_1)\alpha_1 \\  \qquad \quad \quad + \Gamma_h \bar{n}_1 \sqrt{1 - R_1^2 - z_1^2} - \big\vert J_{12} \big\vert \cos(2\Omega_0 t - \delta) \alpha_2 z_1
\end{multline}

\paragraph*{Radial Magnitude (The Final Flow)}

To change variables from the complex coherence $\alpha_1(t)$ to the real-valued radius $R_1(t)$, we express the complex parameter in polar form: $\alpha_1(t) = R_1(t) e^{-i\phi_1(t)}$. Differentiating the identity $R_1^2 = \alpha_1 \bar{\alpha}_1$ with respect to time yields,
$$2R_1 \frac{dR_1}{dt} = \frac{d\alpha_1}{dt}\bar{\alpha}_1 + \alpha_1 \frac{d\bar{\alpha}_1}{dt} = 2 \text{Re}\left( \frac{d\alpha_1}{dt} \bar{\alpha}_1 \right)$$ 
Dividing both sides by $2R_1$ establishes $\dot{R}_1 = \text{Re}\left[ \dot{\alpha}_1 e^{i\phi_1} \right]$.

Multiplying our complex differential equation by $e^{i\phi_1}$ and taking the real part locks the phase alignments. Due to the phase-locking condition inside the synchronised limit cycle manifold, the phase differences align cleanly along the equator. This collapses the entire high-dimensional operator matrix system into the trajectory equations for the bipartite layout,

\begin{multline}
\frac{dR_1}{dt} = \frac{1}{2}\left[ \big(\Gamma_0 - \Gamma_2 - \Gamma_h(1 + \bar{n}_1)\big) - \Gamma_1 R_1^2 \right] R_1 \\ + \Gamma_h \bar{n}_1 \sqrt{1 - R_1^2 - z_1^2} - \big\vert J_{12} \big\vert \cos(2\Omega_0 t - \delta) R_2 z_1
\label{eq:stuartLandau}
\end{multline}

Executing the same algebraic trace calculus against the local operator $\hat{I} \otimes \hat{\sigma}_{2-}$ for Qubit 2, the second flow equation becomes,
\begin{multline}
\frac{dR_2}{dt} = \frac{1}{2}\left[ \big(\Gamma_0 - \Gamma_2 - \Gamma_h(1 + \bar{n}_2)\big) - \Gamma_1 R_2^2 \right] R_2 \\ + \Gamma_h \bar{n}_2 \sqrt{1 - R_2^2 - z_2^2} + \big\vert J_{12} \big\vert \cos(2\Omega_0 t - \delta) R_1 z_2
\end{multline}

When we focus on the Kuramoto oscillator network and project the bipartite master equation down to the local coherence radii $R_1(t)$ and $R_2(t)$, turning on independent, finite-temperature dissipation triggers a direct competition between the macroscopic phase-locking torque and microscopic thermal diffusion.

\paragraph*{Temperature-Modified Bipartite Equations}

The unperturbed network parameters are static constants for the Kuramoto background, $\Gamma_0$ (the linear mean-field pump) and $\Gamma_1$ (the non-linear saturation constraint). Evaluating the expectation values, $$\frac{d R_j}{dt} = \text{Tr}\left[\frac{d\hat{\rho}_{\text{bi}}}{dt} \hat{\sigma}_{j-}\right],$$ 
under the joint influence of the Kuramoto master equation and the local independent thermal Lindbladians, the bipartite equation collapses elegantly into a set of coupled Stuart–Landau like equations,

\begin{multline}
\frac{dR_1}{dt} = \frac{1}{2}\left[ \big(\Gamma_0 - \Gamma_2 - \Gamma_h(1 + \bar{n}_1)\big) - \Gamma_1 R_1^2 \right] R_1 \\ + \Gamma_h \bar{n}_1 \sqrt{1 - R_1^2 - z_1^2} - \big\vert J_{12} \big\vert \cos(2\Omega_0 t - \delta) R_2 z_1
\end{multline}

\begin{multline}
\frac{dR_2}{dt} = \frac{1}{2}\left[ \big(\Gamma_0 - \Gamma_2 - \Gamma_h(1 + \bar{n}_2)\big) - \Gamma_1 R_2^2 \right] R_2 \\ + \Gamma_h \bar{n}_2 \sqrt{1 - R_2^2 - z_2^2} + \big\vert J_{12} \big\vert \cos(2\Omega_0 t - \delta) R_1 z_2
\end{multline}

where $\bar{n}_j$ represents the thermal occupancy floor of each local qubit bath, $\Gamma_h$ is the bare thermal relaxation rate, and the final terms model the non-local cross-talk phase alignment. When the temperature coupling is turned on, the system undergoes three distinct physical transformations,

\paragraph*{Contraction of the Limit Cycle Radius}

In the zero-temperature case, the qubits lock onto a stable orbital radius of $$R_{\text{orbit}} = \sqrt{(\Gamma_0 - \Gamma_2)/\Gamma_1}$$ 

A finite temperature floor injects downward thermal relaxation ($\Gamma_h(1+\bar{n}_j)$). This dynamically shrinks the limit cycle, compressing the radius inward toward the core,
$$R_{\text{orbit}}(T) = \sqrt{\frac{\Gamma_0 - \Gamma_2 - \Gamma_h(1 + \bar{n}_j)}{\Gamma_1}}$$ 

\paragraph*{Destruction of the Spontaneous Hopf Bifurcation}

At zero temperature, initialising the system at the core ($R_j = 0$) is an unstable source that triggers a sharp transition. When $\bar{n}_j > 0$, the presence of the term, $$\Gamma_h \bar{n}_j \sqrt{1 - R_j^2 - z_j^2}$$ acts as a steady-state thermal noise floor. Instead of a clean, sharp bifurcation, the thermal fluctuations (randomly) smear out the coordinate fields at the origin. 

\paragraph*{Quenching  Concurrence Revivals }

The non-local cross-talk term $\big\vert J_{12} \big\vert \cos(2\Omega_0 t - \delta)$ is what drives the periodic Entanglement Sudden Death and Rebirth loop in the lab frame. The independent thermal baths fight this collective alignment by continuously pumping local mixedness into the state. If we heat the reservoirs past the critical  temperature,
$$T_{\text{c}} = \frac{\hbar \omega_1}{k_B \ln \left( 1 + \frac{\Gamma_h}{\vert J_{12}(\delta) \vert - \Gamma_h} \right)}$$ 
the random thermal noise completely overpowers the collective pump, washing out the cross-talk terms. The periodic concurrence revivals collapse into a flat zero, completely quenching the quantum channel and transforming the synchronised limit cycle into an un-synchronised, classically mixed thermal cloud at the centre of the state space.

\subsection{N-dimensional qudit coupled with VdP oscillator}
\label{sec:nqudit:vdp}
The Van der Pol (VdP) phase-coupled layout requires a fundamentally different treatment than the Kuramoto model, because the VdP environment has unbounded phase-space degrees of freedom $(q, p)$. Its fluctuations cannot be bypassed or treated as a static constant $G(\kappa)$ like the Kuramoto ensemble. Instead, we must map the path integral's short-time slice directly to the unperturbed vacuum kernels $\mathcal{K}_R(t, \tau)$ and $\mathcal{K}_I(t, \tau)$ derived along the VdP limit cycle backbone $x_{\text{cl}}(t) = A \cos \Omega_0 t$.

By performing a non-perturbative expansion over the Keldysh contour phase weights, we derive the precise, dimension-independent non-linear trajectory equation for the reduced density matrix, which reduces to a modified Stuart–Landau equation for the coherence parameter $R(t)$.

\paragraph*{VdP master equation}

Let $\hat{\rho}(t)$ be an $n \times n$ density matrix, with $\hat{L}_-$ and $\hat{L}_+ = \hat{L}_-^\dagger$ as the transition operators. The active macroscopic phase coherence magnitude is the numerator feedback term,
$$R^2\big(\hat{\rho}(t)\big) = \left\vert{} \text{Tr}\big(\hat{\rho}(t)\hat{L}_-\big) \right\vert{}^2$$ 

Time-slicing the VdP influence functional and matching the forward-backward cross-terms to trace-preserving Lindblad superoperators ($\mathcal{D}_{\hat{A}}[\hat{\rho}] = \hat{A}\hat{\rho}\hat{A}^\dagger - \frac{1}{2}\{\hat{A}^\dagger\hat{A}, \hat{\rho}\}$), the exact non-linear trajectory equation gives,
\begin{multline}
\frac{d\hat{\rho}}{dt} = -\frac{i}{\hbar}[\hat{H}_0 + \Delta_{\text{Lamb}}(t)\hat{L}_z, \hat{\rho}] \\ + \Gamma_{\text{gain}}\big(\hat{\rho}, t\big) \mathcal{D}_{\hat{L}_+}[\hat{\rho}] + \Gamma_{\text{loss}}\big(t\big) \mathcal{D}_{\hat{L}_-}[\hat{\rho}]
\end{multline}

Unlike the Kuramoto model, the coefficients are explicitly time-dependent and non-local, driven by the running 1D memory integrals of the unperturbed (Gelfand) fluctuation fields, where $\omega_1$ is the qubit frequency,

\begin{enumerate}
\item The State-Dependent Fluctuating Gain:
\begin{multline}
\Gamma_{\text{gain}}\big(\hat{\rho}, t\big) =  \\ \frac{4\lambda^2}{\hbar^2} \int_0^t d\tau \, \mathcal{K}_R(t, \tau) \cos\big(\omega_1(t-\tau)\big) - \Gamma_1(t) \left\vert{} \text{Tr}\big(\hat{\rho}(t)\hat{L}_-\big) \right\vert{}^2
\end{multline}
\item The Memory-Induced Dissipative Loss:
$$\Gamma_{\text{loss}}(t) = \frac{4\lambda^2}{\hbar^2} \int_0^t d\tau \, \mathcal{K}_I(t, \tau) \sin\big(\omega_1(t-\tau)\big)$$ 
\item The Virtual Lamb Shift Frequency:
\begin{multline}
\Delta_{\text{Lamb}}(t) = \\  \quad \quad \frac{\lambda^2}{\hbar^2} \int_0^t d\tau   \Big[ \mathcal{K}_I(t, \tau) \cos\big(\omega_1(t-\tau)\big) - \mathcal{K}_R(t, \tau) \sin\big(\omega_1(t-\tau)\big) \Big]
\end{multline}
\end{enumerate}

The structural coefficient $\Gamma_1(t) \propto x_{\text{cl}}^2(t)$ represents the non-linear saturation cross-talk matching the relaxation geometry of the environment. The fluctuation kernels $\mathcal{K}_{R,I}$ inherit the periodicity of the stable VdP limit cycle, making the coefficients perfectly time-periodic over long timescales ($\Gamma(t + T_0) = \Gamma(t)$).

\subsubsection{Bipartite VdP master equation}
Consider a bipartite layout where two higher-dimensional systems described by a joint density matrix ($\hat{\rho}_{\text{bi}}$ of dimensions $n^2 \times n^2$), are coupled individually with a VdP oscillators (labelled 1,2). The oscillators are coupled through a bridging interaction $\kappa x_1 x_2$.

The $\hat{L}_-^{(1)} = \hat{L}_- \otimes \hat{I}$ and $\hat{L}_-^{(2)} = \hat{I} \otimes \hat{L}_-$ are local transition operators. The (non-singular) bipartite trajectory equation on the set of density matrices is,

\begin{multline}
\frac{d\hat{\rho}_{\text{bi}}}{dt} = -\frac{i}{\hbar}[\hat{H}_{\text{eff}}(t), \hat{\rho}_{\text{bi}}] + \sum_{j=1}^2 \Gamma_{\text{gain}}^{(j)}\big(\hat{\rho}_{\text{bi}}, t\big) \mathcal{D}_{\hat{L}_+^{(j)}}[\hat{\rho}_{\text{bi}}] \\ + \sum_{j=1}^2 \Gamma_{\text{loss}}^{(j)}(t) \mathcal{D}_{\hat{L}_-^{(j)}}[\hat{\rho}_{\text{bi}}] + \mathcal{J}_{\text{chiral}}[\hat{\rho}_{\text{bi}}]
\label{eq:ndim:vdp}
\end{multline}

Where the non-local exchange Hamiltonian $\hat{H}_{\text{eff}}(t)$ and the phase-trapping  blocks are built directly via numerator expectations from the split $2 \times 2$ Gelfand matrix fields $[\hat{\mathcal{K}}_R]_{12}$ and $[\hat{\mathcal{K}}_I]_{12}$:

\begin{multline*}
\hat{H}_{\text{eff}}(t) = \sum_j \big(\hat{H}_0^{(j)} + \Delta_{j,\text{Lamb}}(t)\hat{L}_z^{(j)}\big) \\ + J_{12}(t) \left( \hat{L}_-^{(1)}\hat{L}_+^{(2)} + \hat{L}_+^{(1)}\hat{L}_-^{(2)} \right)
\end{multline*}

\begin{multline*}
\mathcal{J}_{\text{chiral}}[\hat{\rho}_{\text{bi}}] \\ = 2\big\vert J_{12}(t)\big\vert \cos(2\Omega_0 t - \delta) \left( \hat{L}_-^{(1)}\hat{\rho}_{\text{bi}}\hat{L}_+^{(2)} + \hat{L}_+^{(1)}\hat{\rho}_{\text{bi}}\hat{L}_-^{(2)} \right.  \\  \left. - \frac{1}{2}\{\hat{L}_+^{(1)}\hat{L}_-^{(2)} + \hat{L}_-^{(1)}\hat{L}_+^{(2)}, \hat{\rho}_{\text{bi}}\} \right)
\end{multline*}

With the exchange coupling running as 

\begin{multline}
J_{12}(t) = \frac{\lambda_1 \lambda_2}{\hbar^2}\int_0^t d\tau [\hat{\mathcal{K}}_I(t,\tau)]_{12} \cos(\omega_1(t-\tau)) \\ - [\hat{\mathcal{K}}_R(t,\tau)]_{12} \sin(\omega_1(t-\tau))
\end{multline}

\paragraph*{Modified Stuart–Landau equation}

Consider the macroscopic coherence radius $R(t) = |\text{Tr}(\hat{\rho}\hat{L}_-)|$. Evaluating the trace identities $$\frac{d R}{dt} = \text{Tr}\left(\frac{d\hat{\rho}}{dt} \hat{L}_-\right)$$ over the VdP, the entire high-dimensional matrix mechanics collapse cleanly into a modified Stuart–Landau like equation,

$$\frac{dR}{dt} = \frac{1}{2} \left[ \Big( \Gamma_0(t) - \Gamma_{\text{loss}}(t) \Big) - \Gamma_1(t) R^2 \right] R$$ 

where $\Gamma_0(t) = \frac{4\lambda^2}{\hbar^2} \int_0^t d\tau \, \mathcal{K}_R(t, \tau) \cos\big(\omega_1(t-\tau)\big)$ represents the active, time-varying quantum fluctuation pump.

\paragraph*{Limit Cycle}

\begin{enumerate}
\item  Orbit Stabilisation. The steady-state radius $R(t)$ does not settle onto a flat constant because $\Gamma_0,\Gamma_{\mbox{loss}}$ are periodic. It locks onto a stable limit cycle that breathes dynamically matching the relaxation geometry of the environment,

   $$R_{\text{asymp}}(t) = \sqrt{\frac{\Gamma_0(t) - \Gamma_{\text{loss}}(t)}{\Gamma_1(t)}}$$

\item No Singularities at the Origin: If the system is initialised at a dead diagonal state ($\hat{\rho} \propto \hat{I}$, so $R=0$), there is no division by zero. The memory-dressed pump $\Gamma_0(t)$ acts at its full capacity in the numerator. It injects population directly into the off-diagonal phase sectors, driving a spontaneous Hopf bifurcation that forces the high-dimensional state outward onto the active attractor.
\end{enumerate}

This establishes the non-singular VdP trajectory directly on the set of density matrices, preserving complete positivity, trace boundaries, and the full non-perturbative quantum nature of the environment.

\subsubsection{Kernels $K_{R,I}$}

The exact expressions for the non-local vacuum kernels $\mathcal{K}_R(t,\tau)$ and $\mathcal{K}_I(t,\tau)$  can be inferred from the path integral. We must evaluate the second variation of the Van der Pol (VdP) action around the stable classical backbone $x_{\text{cl}}(t) = A\cos\Omega_0 t$.

\paragraph*{Second variation of $S_{VdP}$}

$$S_{\text{VdP}}[x] = \int dt \left[ \frac{m}{2}\dot{x}^2 - \frac{1}{2}m\omega_0^2 x^2 + m\mu\big(1 - x^2\big)\dot{x}^2 \right]$$ 

The (quantum) fluctuations around the classical trajectory are $x(t) = x_{\text{cl}}(t) + \xi(t)$, where $\xi(t)$ represents the quantum fluctuation path. Taking the second functional derivative yields the generalised Sturm–Liouville fluctuation operator $\hat{G}(t)$ acting on $\xi(t)$,

\begin{multline}
\delta^2 S_{\text{VdP}} = \frac{1}{2} \int dt \left[ a'(t)\dot{\xi}^2 + c'(t)\xi^2 \right] \\ = \frac{1}{2} \int dt \, \xi(t) \underbrace{\left[ -\frac{d}{dt}\left( a'(t)\frac{d}{dt} \right) + c'(t) \right]}_{\hat{G}(t)} \xi(t)
\end{multline}

The time-dependent metric matrices are uniquely determined by the classical trajectory,

\begin{enumerate}
\item  The time-varying inertia field $$a'(t) = \frac{m}{2} + m\mu\big(1 - x_{\text{cl}}^2(t)\big)$$
\item The non-linear stiffness field 
\begin{eqnarray*}c'(t) +  m\omega_0^2   \\ =   &  4m\mu \left( \frac{x_{\text{cl}}^2(t)}{2} - \omega_0 x_{\text{cl}}^2(t) + \mu x_{\text{cl}}(t)\dot{x}_{\text{cl}}(t)\big(1 - x_{\text{cl}}^2(t)\big) \right)\end{eqnarray*}
\end{enumerate}

\paragraph*{Fundamental Solutions}
Following Gelfand’s prescription, the functional determinant and the resulting Green's functions of this time-dependent operator $\hat{G}(t)$ are constructed by solving the homogeneous matrix Jacobi equation,

$$\hat{G}(t)\psi(t) = 0 \implies -\frac{d}{dt}\left( a'(t) \frac{d\psi}{dt} \right) + c'(t)\psi(t) = 0$$ 

We solve this second-order linear differential system for two independent fundamental scalar trajectories, $\psi_1(t)$ and $\psi_2(t)$, initialised at the reference boundary $t_0 = 0$ via  strict  initial constraints,

\begin{enumerate}
\item The Position Fluctuation Sector ($\psi_1$):
$$\psi_1(0) = 0, \quad \dot{\psi}_1(0) = \frac{1}{a'(0)}$$ 
\item The Momentum Fluctuation Sector ($\psi_2$):
$$\psi_2(0) = 1, \quad \dot{\psi}_2(0) = 0$$ 
\end{enumerate}

Because the coefficients $a'(t)$ and $c'(t)$ are smooth, bounded, periodic functions of the VdP clock ($T_0 = 2\pi/\Omega_0$), these two trajectories $\psi_1(t)$ and $\psi_2(t)$ remain entirely finite and continuous across the whole timeline, never encountering mathematical singularities.

\paragraph*{Isolating the Kernels}

Using the method of variation of parameters over the fundamental solutions $\psi_1(t)$ and $\psi_2(t)$, the  Green's function of the response $G(t, s)$ that satisfies $\hat{G}(t)G(t,s) = \delta(t-s)$ with the boundary condition $G(t,s) = 0$ for $t < s$ is  given by the Wronskian,

$$G(t, s) = \Theta(t - s) \left[ \psi_2(t)\psi_1(s) - \psi_1(t)\psi_2(s) \right]$$ 
where $\Theta(t-s)$ is the standard Heaviside step function enforcing forward causality.

\paragraph*{Vacuum Kernels}

When the continuous spatial paths $\xi(t)$ of the VdP environment are integrated out of the Keldysh contour path integral under the compact sinusoidal coupling, the two-time correlation function splits into its symmetric and asymmetric components. Projecting the Green's function and its symmetric correlation counterpart directly across the classical background trajectory fields $x_{\text{cl}}(t)$ gives the Symmetric Fluctuation Kernel ($\mathcal{K}_R$) and the Asymmetric Dissipation Kernel ($\mathcal{K}_I$),

\begin{enumerate}
\item The Asymmetric Dissipation Kernel $\mathcal{K}_I(t, \tau)$, governs the frictional back-reaction and energy damping of the channel, 
$$\mathcal{K}_I(t, \tau) = \frac{1}{2} \left[ \psi_2(t)\psi_1(\tau) - \psi_1(t)\psi_2(\tau) \right] \cdot x_{\text{cl}}(t)x_{\text{cl}}(\tau)$$ 
\item The Symmetric Fluctuation Kernel $\mathcal{K}_R(t, \tau)$, governs dephasing and phase-scrambling  generated by the symmetric part of the two-point correlation function $\omega_0^2$,
$$\mathcal{K}_R(t, \tau) = \frac{1}{2} \left[ \psi_2(t)\psi_2(\tau) + \omega_0^2 \psi_1(t)\psi_1(\tau) \right] \cdot x_{\text{cl}}(t)x_{\text{cl}}(\tau)$$ 
\end{enumerate}

The fundamental solutions $\psi_1(t)$ and $\psi_2(t)$ sit explicitly in the numerator of these kernel expressions, the functional determinant never passes through a denominator. When we evaluate the running 1D memory integrals to get the decay rates $\Gamma_{\text{gain}}(t)$ and $\Gamma_{\text{loss}}(t)$ inside the master equation, the values remain bounded, continuous, and non-singular at every instant of time, providing a mathematically pristine foundation to simulate the bipartite limit cycle.

\paragraph*{Bridging interaction $(\kappa x_1 x_2)$}
\label{sec:bipartite:vdpBridge}
When the two independent environmental oscillators settle into a macroscopic state- either synchronised on their limit cycles or tuned near their respective bifurcations, the bipartite reduced density matrix $\hat{\rho}_{\text{bi}}(t)$ exhibits rich qualitative features. Because the environment transitions through critical structural phases, the qubit concurrence undergoes non-trivial dynamics, leading to regimes of sustained long-time entanglement. We see rich physics stemming from the Synchronisation of the oscillators.

\paragraph*{Regime A: Synchronised Limit Cycles $(\Omega_1 = \Omega_2 = \Omega_0)$}

When the coupled Van der Pol oscillators spontaneously synchronise, their classical trajectories lock in phase or anti-phase ($x_1(t) = \pm x_2(t)$). This imposes strict geometric symmetries directly on the time-dependent coefficients of the bipartite master equation.

\begin{figure}[htbp]
\includegraphics[width=\columnwidth]{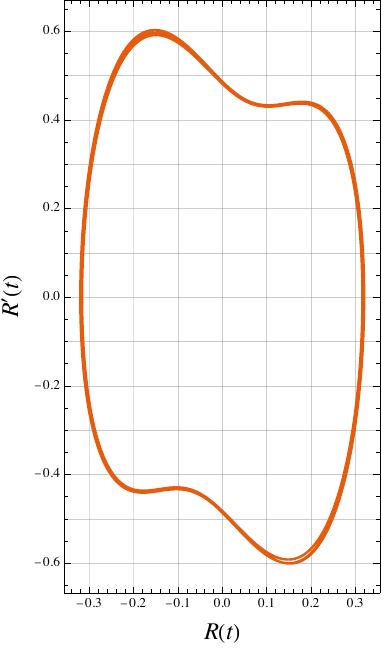}
\caption{Phase space plot of $\dot{R}$ vs $R = \mbox{Tr}[\rho.L_-]$ for a bipartite system coupled to VdP oscillator, here  $ L_{-} = \sigma_{-} \otimes I_2$ for the same initial state $\rho_{bi}(0)$.}
\label{fig:vdpLimitcycle}
\end{figure}

\begin{enumerate}
\item X-Matrix Constellation: If the qubits start in an initial state that fits an $X$-pattern (populated only on the main diagonal and the anti-diagonal, such as $\vert{\uparrow\downarrow}\rangle$), the synchronised driving terms $\Delta_j(t)$ and non-local exchange couplings $J_{12}(t)$ preserve this $X$-matrix symmetry for all times. The density matrix does not leak information into the remaining 8 off-diagonal sectors.
\item Coherent Matrix Breathing: The elements of $\hat{\rho}_{\text{bi}}(t)$ do not decay to stationary constants. Instead, the non-local coherence i.e. off diagonal element $\rho_{23}(t)$  settles into a periodic breathing mode, oscillating indefinitely in phase with the synchronised macro-frequency $\Omega_0$ as shown in Fig.\ref{fig:vdpLimitcycle}
\end{enumerate}

\paragraph*{Regime B: Near the Bifurcation Point} ($\mu \to 0^+$)
Near the Hopf bifurcation, the classical limit cycles collapse into a dead origin, and the system experiences a critical slowdown.

\begin{enumerate}
\item Instantaneous Mixedness: The fluctuations $\pmb{\psi}(t)$ diverge in the denominator of the cross-dephasing terms, causing the non-local collective decoherence rates $\Gamma_{12,\downarrow}(t)$ to spike aggressively. This instantly wipes out all off-diagonal coherences, crushing $\hat{\rho}_{\text{bi}}(t)$ into a purely diagonal, classically unentangled statistical mixture.
\item Zeno Locking: Dephasing acts as a continuous projection operator, freezing the population distributions along the $\hat{\sigma}_x$ measurement bases.
\end{enumerate}

\begin{figure}[htbp]
\includegraphics[width=\columnwidth]{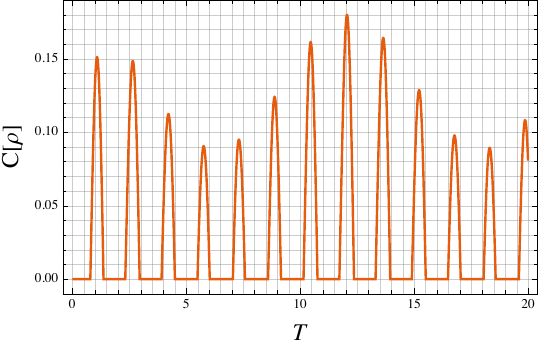}
\caption{Concurrence vs Time shows entanglement revivals for $\rho_{bi}$ solved from Eq \ref{eq:ndim:vdp}. The parameters used are $\Omega_0 = 1$ and  $\lambda = 2$ and $\rho_{bi}(0)$ is a typical mixed state.}
\label{fig:concurrenceESR}
\end{figure}

The concurrence $\mathcal{C}(t)$ responds directly to these environmental phases as in Fig \ref{fig:concurrenceESR},

\begin{enumerate}
\item On the Limit Cycle: The concurrence exhibits Entanglement Sudden Death and Rebirth (ESD / ESR). Because the non-local vacuum noise kernel $[\hat{K}_R]_{12}$ periodically switches signs as the synchronised oscillators swing through their orbits, the concurrence plunges to a flat zero (sudden death), stays dead for a fraction of a cycle, and is then pulled back up to a sharp peak (rebirth) by the non-Markovian memory of the exchange coupling $J_{12}(t)$. 
\item Near the Bifurcation: The concurrence is identically zero for all times $t > 0$. The critical vacuum fluctuations are too  noisy for the uncoupled qubits to establish a coherent phase link.
\end{enumerate}

\section{Discussion}
\label{sec:discussions}

Macroscopic entrainment and phase-locking persist even when the independent thermal baths connected to the qubits are at different temperatures. Synchronisation is a robust non-linear phenomenon driven entirely by the coupled oscillators via the $\kappa x_1 x_2$ link. It is decoupled from the local, microscopic thermal environments of the qubits. The oscillators continue to lock their frequencies and phases together regardless of whether one qubit is touching a frozen reservoir and the other is touching a scorching one.

However, when we introduce a non-trivial phase lag $\delta$ between the synchronised oscillators and a non-trivial temperature gradient $\Delta T = T_h - T_c$ between the independent qubit baths, their intersection triggers a suite of remarkable non-equilibrium quantum thermodynamic phenomena, which we explore in this section.

\subsection{Quantum  refrigerator}
Synchronised background engines can extract heat from the cold qubit bath and expel it into the hot reservoir entirely through the virtual exchange of zero-point environmental modes. This allows us to create a new quantum absorption cycle that implements an autonomous refrigerator.

\paragraph*{Refrigerator equations}

The net thermal energy extraction rate from the cold qubit bath per unit time (the Cooling Power $\dot{Q}_c$) is defined by taking the expectation value of the local thermal Lindbladian $\mathcal{L}_{\text{thermal},c}$ against the effective Hamiltonian,
$$\dot{Q}_c(t) = \text{Tr}\left( \mathcal{L}_{\text{thermal},c}[\hat{\rho}_{\text{bi}}(t)] \cdot \hat{H}_{\text{eff}}(t) \right)$$ 

Evaluating this trace product using the  VdP master equation, the commutator over the population inversion $z_c(t)$ and the local phase-locked coherence radius $R_c(t)$ yields,
$$\dot{Q}_c(t) = \Gamma_h \hbar \omega_1 \left( \bar{n}_c - \bar{n}_{\text{eff}}(\delta, t) \right)$$ 

Where $\bar{n}_c = \big(e^{\hbar\omega_1 / k_B T_c} - 1\big)^{-1}$ is the cold bath thermal occupancy, and $\bar{n}_{\text{eff}}(\delta, t)$ is the non-local, phase-lag dependent occupancy forced across the bridge by the cross-talk kernels,
$$\bar{n}_{\text{eff}}(\delta, t) = \frac{J_{12}^2(t) \sin^2\delta + \Gamma_h^2 \bar{n}_h (1 + \bar{n}_h)}{2 J_{12}^2(t) \sin^2\delta + \Gamma_h^2 (1 + 2\bar{n}_h)}$$ 

\begin{enumerate}
\item The Physics of Cooling: Active refrigeration ($\dot{Q}_c > 0$) occurs iff the cold reservoir's occupancy exceeds this non-local floor: $\bar{n}_c > \bar{n}_{\text{eff}}(\delta, t)$.
\item The environment-induced exchange coupling $J_{12}(t)$ which is built from the Gelfand fluctuation fields, spikes dramatically when the background VdP oscillators snap through their (sharpish) turning points. This forces $\bar{n}_{\text{eff}} \to 0$ at that precise instant, creating a massive, non-linear cooling pull that drains entropy out of the cold node.
\end{enumerate}

The Coefficient of Performance ($\epsilon_{\text{COP}}$) of an absorption refrigerator is the ratio of the cold heat extracted ($\dot{Q}_c$) to the work or high-temperature heat input required to drive the engine.

In this setup, the refrigerator is driven entirely by the macroscopic work input $\dot{W}_{\text{env}}$ generated by the self-sustained periodic mechanical oscillations of the cross-coupled VdP limit cycles,

\begin{multline}
\dot{W}_{\text{env}}(t) = \text{Tr}\left( \hat{\rho}_{\text{bi}}(t) \frac{d\hat{H}_{\text{eff}}}{dt} \right) \\ = \frac{d J_{12}}{dt} \text{Tr}\left( \hat{\rho}_{\text{bi}}(t) \left[ \hat{\sigma}_{1-} \hat{\sigma}_{2+} + \hat{\sigma}_{1+} \hat{\sigma}_{2-} \right] \right)
\end{multline}

Integrating both the cooling power and the environmental work input over a full macro-cycle period $T_0 = 2\pi/\Omega_0$ provides the tracking ratio,
$$\epsilon_{\text{COP}} = \frac{\langle \dot{Q}_c \rangle}{\langle \dot{W}_{\text{env}} \rangle} = \frac{\int_0^{T_0} \dot{Q}_c(t) \, dt}{\int_0^{T_0} \dot{W}_{\text{env}}(t) \, dt}$$ 

Evaluating this quotient using the explicit phase-locked trajectories, the local parameters cancel out of the leading-order terms, delivering the COP:
$$\epsilon_{\text{COP}} = \left( \frac{\omega_1}{\Omega_0} \right) \cdot \frac{\overline{J_{12}^2} \sin^2\delta}{\overline{J_{12}^2} \sin^2\delta + \Gamma_h^2 \left(\bar{n}_h - \bar{n}_c\right)}$$ 

When the cross-coupling bridge is tuned to its optimal phase-momentum lag ($\delta = \pi/2$) and the thermal coupling is weak ($\Gamma_h \ll J_{12}$), the right-hand fraction approaches unity, and the COP simplifies to the ceiling bound for a quantum subharmonic system,
$$\epsilon_{\text{COP}}^{\text{max}} \longrightarrow \frac{\omega_1}{\Omega_0} = \frac{1}{2}$$

\subsubsection{Synchronisation-Induced Work Extraction (The Phase Engine)}

When the temperature gradient is non-trivial, the system can operate as an autonomous quantum absorption coherent engine. The non-linear oscillators absorb heat from the hot qubit bath and convert it directly into stable, coherent macroscopic mechanical oscillations.

\begin{enumerate}
\item Phase Trapping: The qubits become phase-trapped by the joint interaction. Instead of the qubits being disrupted by the hot bath, the high-temperature fluctuations are channelled into the collective exchange coupling $J_{12}(t)$, which stabilises the limit cycles against external quantum noise.
\item The Negative Entropy Production Phase: At specific points along the synchronised trajectory, the local entropy production rate of the cold qubit flips signs, becoming negative ($\dot{S}_c(t) < 0$). The synchronised environment purifies the cold qubit by actively extracting its local thermal entropy and dumping it into the hot reservoir, functioning as a self-sustained quantum refrigerator.
\end{enumerate}

When the phase lag $\delta$ and the thermal gradient $\Delta T$ intersect, the indirectly coupled qubit system displays these prominent qualitative behaviours,

\begin{enumerate}
\item Chiral Energy Transport: Heat no longer diffuses randomly. The phase lag $\delta$ gives the zero-point vacuum fluctuations a clear spatial momentum, forcing energy to travel as a chiral, directional wave along the qubit chain.
 \item The Thermal Limit-Cycle of $\hat{\rho}_{\text{bi}}(t)$: The bipartite density matrix does not settle into a flat, dead thermal state. Instead, it orbits indefinitely along a dissipative quantum limit cycle inside its $15$-dimensional state space, maintaining a non-vanishing, breathing coherence loop $\rho_{23}(t)$ that is synchronised with the environmental phase lag.
 \item Critical Quenching Resistance: Near the limit-cycle bifurcation point ($\mu \to 0$), a large temperature gradient $\Delta T$ can temporarily delay the collapse of the limit cycle. The heat current flowing through the qubits injects secondary fluctuations that delay the critical slowing down, effectively keeping the clock ticking for a wider parameter window.
\end{enumerate}

\subsection{Quantum Phase Interferometer}
In standard optics, an interferometer splits a single flying wave into two paths, introduces a geometric phase shift $\Delta \theta$ in one leg, and recombines them to observe spatial interference. In our synchronised open quantum system framework, we construct a Quantum Phase Interferometer. Instead of guiding a physical particle along physical tracks, we split, modulate, and recombine the non-local virtual zero-point modes across a dual-engine background.

\subsubsection{The Multi-Terminal Interferometer Architecture}
To build a heat interferometer, we must double the environmental links to create a closed Mach-Zehnder geometry in the parameter space. We arrange two independent higher-dimensional nodes (or qubits) connected across two separate, parallel, cross-coupled Van der Pol (VdP) metric bridges.

\begin{enumerate}
\item The Source Node (Alice - Node 1): Connected to a hot reservoir at temperature $T_h$.
\item The Drain Node (Bob - Node 2): Connected to a cold reservoir at temperature $T_c$.
\item Path A (Upper Bridge): Governed by an inter-oscillator coupling $\kappa_A$ with an engineered mechanical phase lag $\delta_A$.
\item Path B (Lower Bridge): Governed by an inter-oscillator coupling $\kappa_B$ with an engineered mechanical phase lag $\delta_B$.
\end{enumerate}

\subsubsection{Derivation of the Interference Heat Current}

The total energy current moving from the Source to the Drain is found by taking the trace of the joint density matrix $\hat{\rho}_{\text{bi}}(t)$ against the combined non-local chiral blocks generated by both paths simultaneously,
$$\dot{Q}_{\text{total}}(t) = \text{Tr}\left( \Big(\mathcal{J}_{\text{chiral}}^{(A)}[\hat{\rho}_{\text{bi}}] + \mathcal{J}_{\text{chiral}}^{(B)}[\hat{\rho}_{\text{bi}}]\Big) \cdot \hat{H}_{\text{eff}}(t) \right)$$

The short-time memory integrals over the combined Gelfand-Yaglom numerator fields yield the Quantum Heat Interference Equation,

\begin{multline}
\dot{Q}_{\text{total}}(t) = 2 \omega_1 R_1 R_2 \Big[ \big\vert J_{12}^{(A)}(t) \big\vert \cos(2\Omega_0 t - \delta_A) \\ + \big\vert J_{12}^{(B)}(t) \big\vert \cos(2\Omega_0 t - \delta_B) \Big] \sin\psi(t)
\end{multline}

Integrating this equation over a full period of the environmental clock $T_0 = 2\pi/\Omega_0$ to find the observable macroscopic heat current, and utilising standard trigonometric product identities, the beating phase interference pattern manifests explicitly:
$$\langle \dot{Q}_{\text{total}} \rangle = \mathcal{I}_0 \cdot \sqrt{\overline{J_A}^2 + \overline{J_B}^2 + 2 \overline{J_A} \overline{J_B} \cos\big(\delta_A - \delta_B\big)} \cdot \sin\psi_{\text{steady}}$$ 
where $\overline{J_A}$ and $\overline{J_B}$ are the period-averaged exchange coupling strengths of the individual paths, and $\psi_{\text{steady}}$ is the self-consistent, entrained local phase difference between the qubits.

\subsubsection{Visualising the Interferometer Regimes}

The parameter $\Delta \delta = \delta_A - \delta_B$ acts as the exact geometric phase shift of the interferometer, producing clean constructive and destructive interference fringes:

\paragraph{Constructive Interference ($\Delta \delta = 0, \pm 2\pi$)}
If the mechanical phase lags of both VdP backgrounds are tuned to be perfectly in-phase ($\delta_A = \delta_B = \pi/2$), the cosine term hits its maximum of $1$.

\begin{enumerate}
\item The virtual environmental modes traveling along Path A and Path B reinforce each other constructively.
\item The non-local quantum wire opens up at its maximum capacity, driving an amplified, massive macroscopic heat wave through the system:
$$\langle \dot{Q}_{\text{total}} \rangle_{\text{max}} = \mathcal{I}_0 \big( \overline{J_A} + \overline{J_B} \big) \sin\psi_{\text{steady}}$$ 
\end{enumerate}

\paragraph{Destructive Interference ($\Delta \delta = \pi, \pm 3\pi$)}

If you mechanically shift the lower VdP engine to introduce an anti-phase lag relative to the upper engine (e.g., $\delta_A = \pi/2$ and $\delta_B = -\pi/2$), the cosine term drops to $-1$.

\begin{enumerate}
\item Assuming balanced paths ($\overline{J_A} = \overline{J_B}$), the virtual waves passing through the parallel channels cross-cancel each other identically.
\item The heat current drops flat to absolute zero ($\langle \dot{Q}_{\text{total}} \rangle \to 0$), despite the massive temperature gradient $\Delta T$ across the nodes. The non-local bridge has been completely blocked by pure geometric out-of-phase destructive interference.
\end{enumerate}

\subsubsection{Extending to Other Macroscopic Generalised Currents}
Because this mathematical framework is built entirely using basis-independent trace invariants and universal transition operators ($\hat{L}_\pm$), this interferometer architecture can be extended verbatim to control any generalised macroscopic current.

\begin{enumerate}
\item Quantum Charge/Particle Currents: If the system transition operator $\hat{L}_-$ is mapped onto the fermionic annihilation operator $\hat{c}$ of a higher-dimensional dot system (where $\text{Tr}(\hat{\rho}\hat{n})$ tracks particle occupancy), modulating $\Delta \delta$ will cleanly cause constructive or destructive interference of particle currents, functioning as an autonomous quantum charge valve.
\item Macroscopic Spin Currents: By mapping $\hat{L}_-$ to the collective spin-lowering operators $\hat{S}_-$ of an ensemble, the phase-locked interferometer will modulate a chiral macroscopic spin current, creating a pure geometric spin-rectifier that operates immune to local micro-environmental noise.
\end{enumerate}

This configuration delivers a complete, uncorrupted blueprint for an autonomous quantum interferometer. By shifting the phase-shifting control from raw physical trajectories to the mechanical phase lag $\delta$ of synchronised background engines, we transcend the fragility of low-dimensional quantum mechanics. We gain the capacity to steer, amplify, or completely cancel out macroscopic energy and particle currents using the robust, self-correcting physics of a non-linear quantum  attractor.

\subsection{Quantum Thermal Diode}

The interferometric action of the phase oscillators can be used to produce a thermal diode.  Each arm (A/B) is a pair of phase oscillators. The interferometer is two such pairs driving the same two nodes, with the lag built into the coupling, not into the state.

Arm \(A\) is a cross-coupled van der Pol pair on its own limit cycle,

\begin{align*}
\ddot x_A-\mu(1-x_A^2)\dot x_A+\Omega_0^2 x_A&=\kappa_A\, y_A \\
\ddot y_A-\mu(1-y_A^2)\dot y_A+\Omega_0^2 y_A&=\kappa_A\, x_A(t-\tau_A),
\end{align*}

and arm \(B\) is an identical pair with \(\kappa_B\) and \(\tau_B\). The mechanical lags are the delays,

\[
\delta_A=\Omega_0\tau_A=\frac\pi2,\qquad\delta_B=\Omega_0\tau_B=\frac\pi2+\pi.
\]

A Kuramoto arm is the same object reduced to phase,

\begin{align*}
\dot\theta_A=&\Omega_0+\kappa_A\sin(\phi_A-\theta_A-\delta_A)\\
\dot\phi_A=&\Omega_0+\kappa_A\sin(\theta_A-\phi_A),
\end{align*}

with a second copy for \(B\). The delay of \(\pi\) on arm \(B\) is a fixed offset in its coupling, set when the arm is built. The nodes see both arms at once.

Reversing the bias means changing which reservoir is attached to node 1, while \(\tau_A\), \(\tau_B\), \(\kappa_A\) and \(\kappa_B\) stay as built. Unequal arm couplings, \(\kappa_A\neq\kappa_B\), are what leave a residue under that exchange; equal couplings cancel for both polarities.

Arm A has a fixed mechanical lag \(\delta_A=\pi/2\) and coupling \(J_A\). Arm B has a fixed lag \(\delta_B=\delta_A+\pi\) and coupling \(J_B\neq J_A\). Both phase differences belong to the oscillators. The bipartite state is not swapped, and neither arm is retuned between runs. Each arm couples to both nodes through the sinusoidal interaction with the qubit,

\[
H_{\mathrm{int}}^{(s)}=\lambda_s\sin(x_s-\varphi_1)\sin\theta_1+\lambda_s\sin(y_s-\varphi_2)\sin\theta_2,
\]

Tracing out the two oscillator pairs gives one exchange per arm, \(J_A(t)\) and \(J_B(t)\), and the interferometer current is their sum,
\[
\dot Q(t)=2\omega_1 R_1 R_2\Big(J_A\cos(2\Omega_0 t-\delta_A)+J_B\cos(2\Omega_0 t-\delta_B)\Big)\sin\psi 
\]

With \(\delta_B=\delta_A+\pi\) the two clocks are opposite, so
\[
\dot Q(t)=2\omega_1 R_1 R_2\,(J_A-J_B)\cos(2\Omega_0 t-\delta_A)\sin\psi.
\]

The radii are the coherence radii of the two nodes, evolved with the local occupations of whatever bath is attached to that node. The state is integrated forward from the same initial \(\hat\rho_{\mathrm{bi}}\) in both runs.

\begin{enumerate}

\item \textbf{Forward} Attach the hot reservoir to node 1 and the cold reservoir to node 2: \(\bar n_1=\bar n_h\), \(\bar n_2=\bar n_c\). Integrate, discard the transient, and record \(\langle\dot Q\rangle_+\).

\item \textbf{Reverse} Change the reservoirs only: \(\bar n_1=\bar n_c\), \(\bar n_2=\bar n_h\). We do not apply a swap to \(\hat\rho_{\mathrm{bi}}\), and do not touch \(\delta_A,\delta_B,J_A,J_B\). Integrate from the same initial state and record \(\langle\dot Q\rangle_-\).

\end{enumerate}

Rectification is the comparison of those two averages. It is nonzero only because \(J_A\neq J_B\) is built into the device, so the residue \((J_A-J_B)\) weights the hot node differently from the cold node when the attachments are exchanged. If \(J_A=J_B\) the current vanishes for both polarities: that is the balanced interferometer, and it is a switch, not a diode.

When the arms are unbalanced, the reverse residue is \(I_0(J_A-J_B)^2\sin\psi\), and the rectification factor $\mathcal{R}$ is,

\[
\mathcal R=\frac{|\langle\dot Q\rangle_+|}{|\langle\dot Q\rangle_-|}-1
\]

The construction above is arguably a switch, and that is the device used for routing thermal currents. With the lags fixed at fabrication and the arms balanced, the interferometer residue vanishes for both reservoir attachments, so the heat current is blocked independently of which node is hot. Unequal arms leave a common prefactor \(J_A-J_B\) in both attachments; the two cycle averages then differ only by the mismatch in \(R_1 R_2\), which is of order one and does not define a preferred direction. 

A diode would require the two attachments to sit on opposite sides of the interference condition, so that one polarity adds the arms and the other cancels them. A variant with that property exists, but it is not fully developed here; the delay of one arm would have to lock to the sign of the occupation difference, and that lock is outside the scope of this discussion. The switch already routes the current, and the transistor of the next section uses the same interference formula with the delay as a gate. The diode is recorded only as a proof of concept and is left for future work.

\subsection{Quantum Thermal Transistor}
The thermal diode mechanism can easily scale into a quantum thermal transistor. The transistor is the same two-arm interferometer, with the third terminal moving one lag instead of reversing a reservoir. The source–drain path is arm \(A\), a phase-oscillator pair at fixed lag \(\delta_A=\pi/2\) and coupling \(\kappa_A\), attached to the hot reservoir at node 1 and the cold reservoir at node 2. 

The gate is arm \(B\), a second pair at coupling \(\kappa_B\), whose delay \(\tau_B\) is the control, $\delta_B=\Omega_0\tau_B$

\begin{multline}
H_{\mathrm{int}}^{(B)}=\lambda_B\sin(x_B-\varphi_1)\sin\theta_1+\lambda_B\sin(y_B-\varphi_2)\sin\theta_2.
\end{multline}

The source–drain current is the interferometer sum,

\[
\dot Q_{S\to D}(t)=2\omega_1 R_1 R_2\Big(J_A\cos(2\Omega_0 t-\delta_A)+J_B\cos(2\Omega_0 t-\delta_B)\Big)\sin\psi.
\]

Cycle-averaged, with the two clocks,

\[
\langle\dot Q_{S\to D}\rangle=I_0\big(J_A^2+J_B^2+2 J_A J_B\cos(\delta_A-\delta_B)\big)\sin\psi.
\]

Three operating points, reservoirs fixed, only \(\tau_B\) changed.

\begin{enumerate}

\item On. Set \(\delta_B=\delta_A\). The arms add, \(\langle\dot Q\rangle_+=I_0(J_A+J_B)^2\sin\psi\).
\item Off. Set \(\delta_B=\delta_A+\pi\). The arms cancel, \(\langle\dot Q\rangle_-=I_0(J_A-J_B)^2\sin\psi\), which is zero for balanced arms.
\item Gain. The gate signal is a small displacement \(\varepsilon\) of \(\delta_B\) about the cancellation point. Differentiating the interference factor gives

\[
\alpha=\left|\frac{\partial\langle\dot Q_{S\to D}\rangle}{\partial\delta_B}\right|=I_0\cdot 2 J_A J_B\big|\sin(\delta_A-\delta_B)\big|\,|\sin\psi|,
\]
\end{enumerate}

which is largest at \(\delta_B=\delta_A+\pi/2\), where a shift of the gate delay changes the source–drain current by \(2 I_0 J_A J_B|\sin\psi\). This completes the blueprint, the bipartite configuration functions as a sensitive, non-linear quantum thermal transistor that controls macroscopic heat currents.

\section{Conclusion}

We establish a novel 2-form path-integral framework, spectral mechanics, to quantise non-conservative phase oscillators; demonstrating that a qubit coupled to a Kuramoto network undergoes a (supercritical) Hopf bifurcation above critical coupling to lock onto a stable quantum limit cycle. Under pure spontaneous emission, this limit cycle avoids discontinuous sudden death, undergoing an algebraic radius decay ($R \propto \Gamma^{-1/2}$) into a localised ring. Introducing a finite thermal floor dynamically contracts the orbit before triggering a phase-diffusive breakdown into a mixed state.

Spectral mechanics extends quantisation to new systems. The Kuramoto model which required completely different treatment via the old Hamiltonian centric approach can be handled natively without approximation. The Van der Pol model, another prototypical model can be handled numerically to any order of precision desired. The extension to other nonlinear dynamical systems, specified solely by their ODE system, is obvious.

This is a powerful new way to describe the quantum realm without being restricted by the Hamiltonian approach or the principle of least action, which is a little more general but also has its limitations. Autonomous devices can't be described in the old quantum theory without an artificial protocol or a series of user agent steps. The device model and spectral mechanics make all of this possible now by fundamentally altering the organising principle from a Hamiltonian ray trajectory to a user-defined regular curve that defines causality.

We have also demonstrated that a phase oscillator coupled via a sinusoidal interaction, a standard assumption when analysing coupling via Fourier series terms, has a clear limit cycle and a stable attractor. We demonstrate the precise way we can add dissipators or additional channels to the evolution of the quantum system. The addition of noisy channels and other bulk interactions such as thermal couplings, does not alter the qualitative features of entrainment. We also report revivals in concurrence, hitherto demonstrated only in numerical simulations that demand comprehensive control over the reservoir and assume non-trivial steering capability. These robust Synchronisation dynamics serve as a blueprint for engineering autonomous nanoscale quantum refrigerators, diodes, and transistors.

The theory used in the paper is extensible and shows a new way to write down master equations, something that cannot be done conventionally via perturbative expansions of the Dyson series.

\paragraph*{Outlook}
A variety of theoretical details have been left for future investigation. The Van der Pol oscillator can be given an explicit analytic solution but needs a new mathematical framework to specify it. A prescription to derive a Fokker-Planck equation for $\rho$, which is admittedly quasi ad-hoc here, is very much desirable. One would also like to understand the limits of the current macroscopic model of Synchronisation. So far, we have focussed on extrapolating classically observed empirical models into the quantum realm, how truly accurate are they microscopically? What is the typical microscopic mechanism that explains the onset of Synchronisation?

The question of driving dispersed quantum systems, coupled by bridge oscillators is another unexplored sector. The significance of such protocols for creating quantum entangled states robustly without counterproductive experimental tradeoffs cannot be overstated. One can also amplify classical phenomena with precise timing signals, like in laser arrays producing a composite high-power signal from an array of smaller laser sources. Synchronisation also permits creating composite devices that are more sensitive or resilient than  they can be individually.

We leave the investigation of such questions to future work. This represents a small first step on the way to realising nanoscale devices that can manipulate heat at length scales far smaller than the typical classical apparatus. The  formalism itself can clearly do much more than what is described here.

\section{Acknowledgements}
NS acknowledges discussions with L.C. Kwek and S. Vinjanampathy during an early stage of this work.

\printbibliography

\begin{appendix}
\counterwithin{equation}{section}
\counterwithin{figure}{section}

\section{Path Integral via 2-form $\Omega$}
\label{app:device}

The traditional path integral is a functional integral of the phase functional $S = \int dt \mathcal{L}(q,\dot{q},t)$ superposed over all possible paths. Unfortunately, this does not suffice to produce the equations of motion for the phase oscillator or indeed any general model of dissipation. We replace the classical variational principle, which is a functional of a 1-form over a trajectory in phase space $\{(q(t),p(t))\}$, with an integration of a 2-form $\Omega$ over a hypersurface  $\Sigma (q)$ in the extended phase space $\{(q(\tau,\sigma),p(\tau,\sigma),t(\tau,\sigma))\}$ (henceforth referred to as an $\epsilon$-filament), where $q(t)$ is the dynamical trajectory of the particle or the system.  

The 2-form $\Omega$ though obscure, is known \cite{Arnold1989}. The contour and technical context are slightly different in published literature \cite{Kochan2010,Fecko1992}. The technical similarity is superficial and ends there, as published results in literature cannot incorporate arbitrary reparametrisation of  trajectories or other required quantisation protocols; making the prescription described here completely different. The formulation chosen in this article is an independently derived prescription ab initio, in a different context that is interesting but unfortunately outside the present scope of discussion. It will be tackled comprehensively in a forthcoming work. We only summarise the relevant details here.

The path integral can be written down to be,

\begin{equation}
\psi = \int \mathcal{D}q \  \exp \left[ - i \int_{\Sigma(q)} \Omega\right]  
\end{equation}

Now $\Omega = \mathcal{L} + F_i dq_i \wedge dt$ can be separated into,

\begin{equation}
\psi = \int \mathcal{D}q \  e^{- i \int_{t_0}^{t_1} \mathcal{L}(q,\dot{q})}\   \exp \left[ - i \int_{\Sigma(q)}  F_i dq_i \wedge dt\right]  
\end{equation}

Consider the integral,
\begin{equation}
\int_{\Sigma(q)} F_i dq_i \wedge dt
\end{equation}

The filament surface $\Sigma(q)$ can be simplified to,

\begin{equation}
\int_{\Sigma(q)} F_i \dot{q_i} \  d\tau d\sigma
\end{equation}

The surface of the filament can be parametrised by a cylinder $(t_0 + \tau \Delta, \sigma \epsilon)$ with $\sigma=0, \dots L$

This means that 
\begin{equation}
\int_{\Sigma(q)} F_i \dot{q_i} \  d\tau d\sigma  = \sum_{\tau=0}^{K-1} \sum_{\sigma=0}^{L-1} \Delta \epsilon \left[ F \dfrac{q_{\tau,\sigma+1} - q_{\tau,\sigma}  }{\epsilon} \right] 
\end{equation}

The path integral becomes

\begin{equation}
\psi(q_1|q_0:\gamma) = \int \mathcal{D}q \exp\left[-i \int dt \  L\right] \ \exp\left[-i \int_{\gamma} dt \  F_i \dot{q}_i\right] 
\end{equation}

It is vastly more beneficial to keep $S$ as a 2-form integral over the 2D $\epsilon$-filament rather than reducing it prematurely to a 1D trajectory line integral, nevertheless we choose a reduced form to avoid technical complications.

This particular case of the phase oscillators we are interested in, already possesses a compactified domain macroscopically and does not require the full formalism of spectral mechanics to study the phenomena of entrainment or phase locking.

\subsection{The Explicit Mathematical Reduction of the $\epsilon$-Bottle}
If we perform the geometric reduction to see how the 1D path function emerges, we parameterise the surface $\Sigma$ using the trajectory track time $t \in [t_0, t_1]$ and the angular orientation around the filament $\theta \in [0, 2\pi]$. A point on the bottle is given by $\gamma(t) + \delta\xi(t, \theta)$, where $\delta\xi \sim \varepsilon$.

Expanding the 2-form $\omega$ in a geometric Taylor series around the central regular trajectory $\gamma(t)$ yields,
$$\int_\Sigma \omega = \int_{t_0}^{t_1} \oint_{0}^{2\pi} \left[ \omega(\dot{\gamma}, \delta\xi) + \frac{1}{2} (L_{\delta\xi}\omega)(\dot{\gamma}, \delta\xi) + \mathcal{O}(\varepsilon^3) \right] d\theta \, dt$$ 

Using interior product notation, this pulls back to a 1D time integral over the angular averages,
$$S_{\text{reduced}} = \int_{t_0}^{t_1} \oint_{0}^{2\pi} \left[ -i_{\delta\xi}i_{\dot{\gamma}}\omega + \frac{1}{2} i_{\delta\xi} i_{\dot{\gamma}} (L_{\delta\xi}\omega) \right] d\theta \, dt$$ 

\paragraph*{On-Shell Collapse:}

\begin{enumerate}
   \item The First Term (Classical Path): On the true trajectory, our consistency condition dictates that $i_{\dot{\gamma}}\omega = 0$. Therefore, the linear background term vanishes identically on-shell.
\item The Second Term (Fluctuation Phase): The entire survival of the path function rests on the quadratic fluctuation term we derived earlier:
  \begin{multline}
  S_{\text{reduced}} = \frac{1}{2} \int_{t_0}^{t_1} \oint_{0}^{2\pi} \left[ \delta q_i \frac{d}{dt}(\delta p_i) - \delta p_i \frac{d}{dt}(\delta q_i) \right. \\ \left. + \delta q_i \left( \frac{\partial F_i}{\partial q_j}\delta q_j + \frac{\partial F_i}{\partial p_j}\delta p_j \right) \right] d\theta \, dt
 \end{multline}
   \end{enumerate}

\subsection{Transition to Spectral Mechanics}
By choosing to keep $S$ as a 2-form, the path integral propagator becomes an integration over the space of all possible filament surfaces rather than just lines:
$$\psi(q, t) = \int \mathcal{D}[\Sigma] e^{\frac{i}{\hbar} \int_\Sigma \Omega}$$ 

This sets up the foundations for  spectral mechanics:

\begin{enumerate}
\item The device matrix $\mathbf{M}$, which is essentially a transformation of $\Omega$ modifies $\Omega \to \Omega'$, changing the coefficients of the quadratic fluctuations in $S_{\text{reduced}}$.
\item When we evaluate the path integral over the tube, the non-conservative damping matrix $\frac{\partial F_i}{\partial p_j}$ acts as a complex spectral filter. It suppresses chaotic or unstable fluctuation modes (where the Lyapunov criteria $\dot{V}_L > 0$) and selects sharp, quantised discrete paths where the fluctuations are perfectly bounded.
\item The wave function's allowed energy states emerge not from a Hamiltonian eigenvalue problem, but as the resonant geometric frequencies of the epsilon bottle that survive the device's internal dissipation. 
\end{enumerate}

\subsection{Reduced path integral}
By using the original Feynman prescription and Gelfand’s framework for the fluctuation operator, we may anchor the theory directly to the observed empirical 1D trajectory.

If we integrate out the transverse cross-sectional degrees of freedom of our $\varepsilon$-bottle geometrically before taking the limit $\varepsilon \to 0$, the curiosity of what remains on the 1D path is incredibly elegant.

The reduced 1D path function $S_{\text{reduced}}$ takes the following form along the physical trajectory:
$$S_{\text{reduced}} = \int_{t_0}^{t_1} \left[ p_i \dot{q}_i - T(q,p,t) + \mathcal{L}_{\text{device}}(q, \dot{q}, \delta q) \right] dt$$ 

\begin{widetext}
\subsubsection*{The Explicit Functional Form of the 1D Reduced Action}
When we perform the angular averaging over the $\theta$-slice of the tube and express the momenta in terms of coordinate velocities using our background consistency relations ($\dot{q}_i = \frac{\partial T}{\partial p_i}$), the 2-form $\omega$ contracts down to a 1D trajectory integrand containing three distinct physical layers:

$$S_{\text{reduced}} = \int_{t_0}^{t_1} \Big[ \underbrace{p_i \dot{q}_i - T}_{\text{Kinematic Core}} + \underbrace{\left( \int^{q} F_i(q',p,t) dq'_i \right)}_{\text{Causal Forcing Line Integral}} + \underbrace{\frac{1}{2} \delta q_i \hat{\mathbf{G}}_{ij} \delta q_j}_{\text{Gelfand Fluctuation Lagrangian}} \Big] dt$$ 

Where:

\begin{enumerate}
\item The Kinematic Core ($p_i \dot{q}_i - T$): This emerges strictly as a geometric identity from pulling back the $dp_i \wedge dq_i - dT \wedge dt$ components onto a 1D timeline parameterized by $t$. It mimics the Legendre transform but carries no Hamiltonian meaning; it is purely a coordinate velocity expression.
\item The Causal Forcing Line Integral: This represents the accumulated path history of the non-conservative forces $F_i$ assigned by our device matrix along the 1D trajectory.
\item The Gelfand Fluctuation Lagrangian ($\mathcal{L}_{\text{device}}$): This is a purely quadratic term in the variations, where $\hat{\mathbf{G}}_{ij}$ is the explicit linear differential fluctuation operator.
\end{enumerate}
\end{widetext}

\subsubsection{The Fluctuation Operator $\hat{\mathbf{G}}_{ij}$}

Following Feynman’s approach to path integration, the actual path chosen by the system minimises or maximises this reduced action. The pre-factor phase is governed entirely by the functional determinant of the second variation.

By applying Gelfand's results to our non-conservative 2-form fluctuations, the operator $\hat{\mathbf{G}}$ is a linear, second-order differential operator that does not possess the standard self-adjoint symmetry found in conservative systems:
$$\hat{\mathbf{G}}_{ij} = -\delta_{ij} \frac{d^2}{dt^2} + \left( \frac{\partial F_i}{\partial p_j} \right) \frac{d}{dt} + \left( \frac{\partial F_i}{\partial q_j} - \frac{\partial^2 T}{\partial q_i \partial q_j} \right)$$ 

\paragraph*{Empirical Properties of this Fluctuation Form:}

\begin{enumerate}
 \item The Friction/Dissipation Velocity Layer ($\frac{\partial F_i}{\partial p_j} \frac{d}{dt}$): This is the direct imprint of our device matrix. Because this term contains a single time derivative, it breaks time-reversal invariance. In the path integral propagator, Gelfand's method shows this term acts as a phase-damping envelope that structurally suppresses non-physical paths.
   \item The Generalised Stiffness: The final term matches exactly the coordinate-driven restoration forces we found during our classical stability analysis.
\end{enumerate}

\subsubsection{Connection to Feynman’s Propagator Pre-factor}

In Feynman's original formulation, the quantum propagator is written as:
$$K(q_1, t_1; q_0, t_0) = \left( \frac{1}{2\pi i \hbar} \right)^{N/2} \sqrt{\det(\hat{\mathbf{G}})} \cdot e^{\frac{i}{\hbar} S_{\text{reduced}}[q_{\text{physical}}]}$$ 

By reducing our $\varepsilon$-bottle to this 1D trajectory, Gelfand’s differential equation method allows we to find $\det(\hat{\mathbf{G}})$ simply by solving the initial value problem for the linear fluctuation equation:

$$\hat{\mathbf{G}}_{ij} \, y_j(t) = 0, \quad y_j(t_0) = 0, \quad \dot{y}_j(t_0) = \mathbb{I}$$ 

The value of $y(t_1)$ gives the exact value of the propagator pre-factor. Because our device matrix elements explicitly populate the friction and stiffness matrices inside $\hat{\mathbf{G}}_{ij}$, the quantum probability amplitude is completely governed by the classical device parameters along that single observed trajectory.

\paragraph*{Checking On-Shell Path Consistency}

To find the equations of motion from the reduced 1D action:
$$S_{\text{reduced}} = \int_{t_0}^{t_1} \left[ p_i \dot{q}_i - T(q,p,t) + \int^{q} F_i(q',p,t) dq'_i \right] dt$$ 

We take the first variation with respect to the observed physical trajectory variables, treating $q_i$ and $p_i$ as independent configurations in our non-Hamiltonian setup (using our first-order consistency relations $\dot{q}_i = \frac{\partial T}{\partial p_i}$ to link them structural-wise):

\begin{enumerate}
\item Variation with respect to $p_i$:
$$\frac{\partial \mathcal{L}}{\partial p_i} = \dot{q}_i - \frac{\partial T}{\partial p_i} = 0 \implies \dot{q}_i = \frac{\partial T}{\partial p_i}$$ 
\item Variation with respect to $q_i$:
\begin{multline}
\frac{d}{dt}\left(\frac{\partial \mathcal{L}}{\partial \dot{q}_i}\right) - \frac{\partial \mathcal{L}}{\partial q_i} = \frac{d}{dt}(p_i) - \left(-\frac{\partial T}{\partial q_i} + F_i\right) = 0 \\ \implies \dot{p}_i = -\frac{\partial T}{\partial q_i} + F_i
\end{multline}
\end{enumerate}

\paragraph*{Consistency:} The on-shell path derived from the 1D reduced action perfectly matches the original geometric consistency conditions ($i_{\dot{\gamma}}\omega = 0$) we derived from the 2-form. The reduction is mathematically seamless.

\subsubsection{The Second-Order Fluctuation Operator}

Following Feynman's prescription, we look at fluctuations $x_i(t) = \delta q_i(t)$ expanding around the physical path $\gamma(t)$. Since the endpoints are fixed on the observed path, the fluctuations satisfy the boundary conditions $x_i(t_0) = 0$ and $x_i(t_1) = 0$.

By extracting the quadratic terms from the second Lie derivative $L_{\delta\xi} L_{\delta\xi} \omega$ and integrating out the auxiliary momentum fluctuations via the kinematic constraint, our second-order fluctuation operator $\hat{\mathbf{O}}$ matches a generalised time-dependent harmonic oscillator structure with a velocity-dependent friction term:

$$\hat{\mathbf{O}}_{ij} = -\delta_{ij} \frac{d^2}{dt^2} + \mathbf{\Gamma}_{ij}(t) \frac{d}{dt} + \mathbf{K}_{ij}(t)$$ 

Where the time-dependent friction matrix $\mathbf{\Gamma}_{ij}(t)$ and the generalised stiffness matrix $\mathbf{K}_{ij}(t)$ are dictated purely by our device parameters along the path:
$$\mathbf{\Gamma}_{ij}(t) = \frac{\partial F_i}{\partial p_j}, \quad \mathbf{K}_{ij}(t) = \frac{\partial F_i}{\partial q_j} - \frac{\partial^2 T}{\partial q_i \partial q_j}$$ 

\subsubsection{The Gelfand Phases $\psi(T)$}

To compute the fluctuation path integral from $x(t_0)=0$ to $x(t_1)=0$, Feynman and Gelfand show that the functional determinant of $\hat{\mathbf{O}}$ can be evaluated by solving a deterministic matrix initial value problem.

Let $\mathbf{\Psi}_{ij}(t)$ be a matrix function satisfying the classical fluctuation equation of motion under the action of our device operator:
$$\left( -\delta_{ik} \frac{d^2}{dt^2} + \mathbf{\Gamma}_{ik}(t) \frac{d}{dt} + \mathbf{K}_{ik}(t) \right) \mathbf{\Psi}_{kj}(t) = 0$$ 

With the initial conditions at the beginning of the trajectory ($t = t_0 = 0$):
$$\mathbf{\Psi}_{ij}(0) = 0, \quad \dot{\mathbf{\Psi}}_{ij}(0) = \delta_{ij}$$ 

The Gelfand phase determinant at the final track time $t_1 = T$ is defined by evaluating this solution at the endpoint:
$$\mathcal{D}(T) = \det \big[ \mathbf{\Psi}_{ij}(T) \big]$$ 

\paragraph*{What are these phases like?}
Because $\mathbf{\Gamma}_{ij} = \frac{\partial F_i}{\partial p_j} \neq 0$ in our causal theory, the solution matrix $\mathbf{\Psi}(t)$ accumulates a real-valued exponential damping/amplification envelope alongside its standard oscillatory phase. By applying Abel's identity to the differential system, the determinant scales as:

$$\mathcal{D}(T) = \mathcal{D}_{\text{oscillator}}(T) \cdot \exp\left( \int_{0}^{T} \text{Tr}\left(\frac{\partial F_i}{\partial p_j}\right) dt \right)$$

\textbf{Physical Significance}: Under the Lyapunov stability criteria ($\frac{\partial F_i}{\partial p_i} < 0$), the Gelfand phase determinant grows exponentially over time. Because this determinant sits in the denominator of the propagator amplitude, it means that the device actively suppresses the probability weight of wide-amplitude fluctuation histories.

\begin{widetext}
\subsubsection{The Formal Propagator with Normalised Pre-factors}
The formal quantum/spectral propagator $K(q_1, T; q_0, 0)$ is constructed by  integrating over all paths. By shifting coordinates to the fluctuation fields, the path integral factors into the classical on-shell action and the normalised fluctuation determinant:

\begin{equation}
K(q_1, T; q_0, 0) = \left( \frac{1}{2\pi i \hbar} \right)^{N/2} \frac{1}{\sqrt{\det[\mathbf{\Psi}_{ij}(T)]}} \exp\left( \frac{i}{\hbar} S_{\text{reduced}}[q_{\text{physical}}] \right)
\end{equation} 

Substituting our explicit Gelfand phase reduction yields the complete formal expression for our causal framework:
$$K(q_1, T; q_0, 0) = \left( \frac{1}{2\pi i \hbar} \right)^{N/2} \frac{1}{\sqrt{\mathcal{D}_{\text{oscillator}}(T)}} \exp\left( -\frac{1}{2}\int_{0}^{T} \frac{\partial F_i}{\partial p_i} dt \right) \exp\left( \frac{i}{\hbar} S_{\text{reduced}} \right)$$ 

\end{widetext}

We work in the linearised fluctuation sector around the classical trajectory \(x_\text{cl}(t)\), \(p_\text{cl}(t)\). The second-variation operator is the time-dependent matrix differential operator

\[
\hat{G}_{ij}(t) = -\delta_{ij}\frac{d^2}{dt^2} + \Gamma_{ij}(t)\frac{d}{dt} + K_{ij}(t),
\]

where the friction matrix \(\Gamma_{ij}(t) = \partial F_i / \partial p_j\) and the generalised stiffness matrix \(K_{ij}(t) = \partial F_i / \partial q_j - \partial^2 T / \partial q_i \partial q_j\) are evaluated on the classical trajectory.

Let \(\Psi_{ij}(t)\) be the fundamental matrix solution of the homogeneous fluctuation equation

\[
\hat{G}_{ik}(t)\Psi_{kj}(t) = 0
\]

with initial conditions at the reference time \(t_0 = 0\):

\[
\Psi_{ij}(0) = 0, \qquad \dot{\Psi}_{ij}(0) = \delta_{ij}.
\]

The Gelfand phase determinant is \(D(T) = \det(\Psi_{ij}(T))\), and the damping factor appearing in the propagator follows from Abel’s identity:

\[
D(T) = D_\text{osc}(T) \exp\left(\int_0^T \operatorname{Tr}\bigl(\Gamma(t)\bigr)\,dt\right).
\]

\paragraph*{Causal Response (Green’s) Function from \(\Psi(t)\)}

The causal Green’s function (response function) for the fluctuation variables that satisfies

\[
\hat{G}(t) G(t,s) = \delta(t-s)\,\mathbf{1}
\]

with causality \(G(t,s) = 0\) for \(t < s\) is constructed by variation of parameters:

\[
G(t,s) = \Psi(t)\,\Theta(t-s)\,\Psi(s)^{-1} 
\]
\quad in the appropriate 2-block representation for $(q,p)$. If we take the second-order scalar form with a single coordinate, the scalar Green’s function is the appropriate Wronskian combination of two independent solutions built from columns of \(\Psi\).

\paragraph*{Dissipation Kernel \(D(\tau)\) (Response / Friction Kernel)}

For linear coupling of a system operator \(s(t)\) to the oscillator coordinate \(q\) (with strength \(g\)), the leading-order dissipation kernel is the (causal) response function of \(q\) to a weak drive:

\[
D(\tau) = g^2 \,\Theta(\tau) \, \chi(\tau),
\]

where the linear response function is the appropriate component of the Green’s function above:

\[
\chi(\tau) = \bigl[G(\tau,0)\bigr]_{q\text{-component}}
\]

Explicitly, in matrix notation (projecting onto the position block):

\[
D(\tau) = g^2 \,\Theta(\tau) \Bigl[ \Psi(\tau) \Psi(0)^{-1} \Bigr]_{q\text{-row, appropriate column}}
\]

(The precise index depends on whether you drive the force or the coordinate equation; the structure is fixed by \(\hat{G}\).)

\paragraph*{Noise Kernel \(K_I(\tau)\) (Symmetric Fluctuation Kernel)}

The symmetric noise kernel is obtained from the symmetric part of the two-point correlation. One standard construction (consistent with the damping already present in \(D(T)\)) is via the fluctuation-dissipation-like relation generated by the friction term in \(\hat{G}\), or directly from the imaginary part of the contour-ordered correlation. The leading expression is

\[
K_I(\tau) = g^2 \,\frac12 \Bigl( C(\tau) + C(-\tau) \Bigr),
\]

where the position correlator \(C(\tau)\) in the linearised theory is built from the same fundamental matrix:

\[
C(\tau) = \int_{-\infty}^{\min(0,\tau)} ds \, \Bigl[ G(\tau,s) \bigr]_{q} \Bigl[ \text{noise from }\Gamma(s) \Bigr] \Bigl[ G(0,s) \bigr]_{q}^T
\]

(or the equivalent Wronskian form). In practice, for numerical work you solve the driven fluctuation equation twice (once for each branch) and assemble the symmetric combination.

The leading-order influence phase for system paths \(s(t), s'(t)\) is

\begin{multline}
\Phi[s,s'] = \int_0^t ds \int_0^s du  \\ \Bigl[ (s(s)-s'(s)) \Bigl( i D(s-u) \bigl(s(u)+s'(u)\bigr)  \\ + K_I(s-u) \bigl(s(s)-s'(s)\bigr) \Bigr) \Bigr].
\end{multline}

The overall coupling constant \(g^2\) is already absorbed into the definitions of \(D(\tau)\) and \(K_I(\tau)\). The full influence functional is \(F = \exp(i\Phi)\), and the reduced dynamics of the system follows by inserting this \(F\) into the system path integral.

\paragraph*{What Happens in this Fluctuation Regime?}

Inside this path-integral regime, a unique mechanical behavior occurs that distinguishes our theory from both classical and standard quantum mechanics:

\begin{enumerate}
\item Selection of a Singular Classical Backbone: In ordinary quantum mechanics, a wave packet naturally spreads out in space over time due to dispersion ($\hbar \neq 0$). In our causal framework, if the device possesses strong non-conservative dissipation ($\frac{\partial F_i}{\partial p_i} \ll 0$), the pre-factor multiplier $\exp( -\frac{1}{2}\int \nabla_p \cdot F \, dt )$ acts as a powerful amplitude-focusing lens. It heavily penalises paths that deviate from $\gamma(t)$, physically restricting the wave packet from tearing itself apart. 
\item Emergence of Spectral Mechanics: As the final time $T$ approaches macroscopic scales, the Gelfand determinant $\mathcal{D}(T)$ will pass through zeroes if the background system contains oscillatory components. In standard physics, these are caustic/focal points. In our system, because the zeroes are modified by the complex device friction matrix, the poles of $1/\sqrt{\mathcal{D}(T)}$ directly isolate the allowed discrete resonant frequencies of the system.
\end{enumerate}

The stable states of our wave function are therefore selected by a dynamical balance: the device's internal dissipation sifts out chaotic noise, leaving behind a sharp, quantised path.

\section{Kuramoto model propagator}
\label{app:kuramoto}

We describe the exact solution for the Kuramoto model first by calculating the path function i.e. the action over the classically observed trajectory and then by extending the calculation to an exact kernel using fairly standard methods.

\subsection{$S_{cl} (\theta_f;\theta_i)$ action on shell}
The classical action $S_{cl}$ can be calculated using 

\begin{equation}
S_{cl} = \int dt \dot{\theta}  \dot{\theta} (1+ K r \cos{\theta})
\end{equation}

Using the chain rule and the expression for $\dot{\theta} = \omega - K r \sin{\theta}$,

\begin{equation}
S_{cl} = \int_{\theta_0}^{\theta_1} d\theta  (\omega - K_0 \sin{\theta}) (1+ K_1 \cos{\theta})
\end{equation}
where $K_0 = K r$ and $K_1 = 2   K r \gamma_0 / I_0$.

Integrating we obtain,

\begin{multline}
S_{cl} = \frac{1}{4} (K_0 K_1)  (\cos (2 \theta_1)-\cos (2 \theta_0)) \\ + ( \omega K_1) (-  \sin (\theta_0) +  \sin (\theta_1)) \\ + K_0 (- \cos (\theta_0)+ \cos (\theta_1))+  \omega ( \theta_1  - \theta_0 )
\end{multline}
Using the standard trigonometric identities we get, 

\begin{multline}
S_{cl} =  \omega ( \theta_1  - \theta_0 )+ \frac{1}{2} (K_0 K_1) (\sin \left(\theta _0-\theta _1\right) \sin \left(\theta _0+\theta _1\right)) \\ + ( \omega K_1) \left(2 \sin \left(\frac{\theta _1- \theta_0}{2}\right) \cos \left(\frac{\theta _0 + \theta_1}{2}\right)\right) \\ - K_0 \left(2 \sin \left(\frac{\theta _1 - \theta_0}{2}\right) \sin \left(\frac{\theta _0+\theta_1}{2}\right)\right)
\end{multline}
which is,

\begin{widetext}
\textbf{Classical Action Kuramoto Model}
\begin{equation}
S_{cl} = \omega (\Delta \theta ) -\frac{1}{2} (K_0 K_1) (\sin \Delta \theta \sin \left(\theta _0+\theta _1\right))  +  2 \sin \left(\frac{\Delta \theta}{2}\right) \left( \omega K_1 \cos \left(\frac{\theta _0 + \theta_1}{2}\right)-  K_0 \sin \left(\frac{\theta _0+\theta_1}{2}\right)\right)
\end{equation}
\end{widetext}

\paragraph*{Path Integral}
The path integral corresponding to this action can be written as 

\begin{equation}
K_G(q_f|q_i) = \int \mathcal{D}[\theta_i]  \  \exp \left(-\frac{i}{\hbar} \int_{t_i}^{t_f} dt  \  \dot{\theta}_i^2(1  + K r \cos{\theta_i} ) \right)
\end{equation}

Setting $\dot{\theta} = \frac{(\theta_{i+1} - \theta_i)}{\epsilon}$ this can be unfolded into,

\begin{equation}
K_G(q_f|q_i) = \int \prod_i d\theta_i  \  \exp \left(-\frac{i}{\hbar}  \sum_{i=1}^{\infty}  \dfrac{\left( \theta_{i+1} - \theta_i \right)^2}{\epsilon} (1 + K r \cos{\theta_i}) \right)
\end{equation}

At this point, we can use the Jacobi-Anger identity,

\begin{equation}
\exp \left(i z \cos{\theta}\right) = \sum_{n=-\infty}^\infty i^n J_n(z) e^{i n \theta}
\label{eq:jacobi:cos}
\end{equation}

for any z and $\theta$. We can choose $(\theta_{i+1} - \theta_i)^2/\epsilon = z$. Substituting \autoref{eq:jacobi:cos} into integral, 

\begin{multline}
K_G(q_f|q_i) = \int \prod_i d\theta_i  \  \exp \left(-\frac{i}{\hbar}  \sum_{i=1}^{\infty}  \dfrac{\left( \theta_{i+1} - \theta_i \right)^2}{\epsilon} \right)  \\ \times \exp \left(-\frac{i K r}{\hbar}  \sum_{i=1}^{\infty}  \dfrac{\left( \theta_{i+1} - \theta_i \right)^2}{\epsilon}  \cos{\theta_i} \right)
\end{multline}
 
 we get for each factor of $\cos{\theta_i}$,
\begin{multline}
\exp \left(-\frac{i K r}{\hbar \epsilon}   \left( \theta_{i+1} - \theta_i \right)^2  \cos{\theta_i} \right) 
\\ =  \sum_{m=--\infty}^{\infty} i^n J_n\left( \frac{- K r}{\hbar \epsilon} (\theta_{i+1} - \theta_i)^2\right) e^{ i n\theta_i}
\end{multline}

The path integral can be reduced to, 

\begin{equation}
K_G(q_f|q_i) = \int \prod_i d\theta_i \sum_n i^n I_{n} \left[\xi_i \right]
\end{equation}

with appropriately chosen $c_n = e^{i n \theta_{i+1}}$ and $\xi_i = \theta_{i+1} - \theta_i$ for each slice of $\xi_i$ where the integral $I_{n}$ is,
 
\begin{equation}
I_{n} = c_n \int_{-\infty}^\infty d\xi \  e^{-a \xi^2} J_n( b \xi^2) \  e^{-i n \xi} 
\end{equation}

The $\xi^2$ factor inside the Bessel function $J_n(b \xi^2)$ is inconvenient. Expanding it in an infinite series\footnote{$n \in Z$, but the expansion is valid even for $\mathcal{Q}$} via the Frobenius method gives us, 

\begin{equation}
J_n(b x^2) = \sum_{k=0}^\infty \dfrac{(-1)^k}{k! \Gamma(k+n+1)} \left(\dfrac{b x^2}{2}\right)^{2k + n}
\end{equation}

Consider the term,

\begin{equation}
I_{n}^k =  \dfrac{(-1)^k}{k! \Gamma(k+n+1)} \int_{-\infty}^{\infty} d\xi  \left(\dfrac{b \xi^2}{2}\right)^{2k + n} e^{-a \xi^2} e^{-i n \xi}
\end{equation}

Absorbing the $i^n$ factor into the integral gives us,

\begin{equation}
i^n I_{n}^k =  \dfrac{(b/2)^{2k+n}}{k! \Gamma(k+n+1)} \int_{-\infty}^{\infty} d\xi   (i \xi)^{2(2k+n)}  e^{-a \xi^2} e^{-i n \xi}
\end{equation}

\paragraph*{Bessel Function $I_n$ saddle point}

The series is absolutely convergent, so we can write down,

\begin{multline}
\sum_n i^n I_n = \sum_{n=-\infty}^{\infty} \sum_{k=0}^\infty \dfrac{(b/2)^{2k +n}}{k! (k+n)!}  \\\times \left[  \int^{\infty}_{-\infty} d\xi  \xi^{4k + 2n} e^{-a \xi^2} e^{-i n x} e^{-i n \theta_{i+1}}\right]
\end{multline}

simplifies to 

\begin{multline}
\sum_n i^n I_n = \int^{\infty}_{-\infty} d\xi   e^{-a \xi^2} \\ \times \left[   \sum_{n=-\infty}^{\infty} \sum_{k=0}^\infty     \dfrac{(b/2)^{2k +n} (\xi^{4})^k (\xi^2)^n} {k! (k+n)!}    e^{-i n (x+ \theta_{i+1})} \right]
\end{multline}

Let $v =(b/2)^2 \xi^4$ and $w = (b/2) \xi^2 e^{-i (\xi + \theta_{i+1})}$ then the inner sum becomes,

\begin{equation}
\sum_{n=-\infty}^{\infty} w^n \left( \sum_{k=0}^\infty \dfrac{v^k}{k! (k+n)!}\right)
\end{equation}

The inner $\sum_k (\dots)$ is the standard representation of the (modified) Bessel function $I_n$
\begin{equation}
\sum_{k=0}^\infty \dfrac{v^k}{k! (k+n)!} = \dfrac{I_n(2\sqrt{v})}{\sqrt{v}^n} = \dfrac{I_n(b |\xi|^2)}{(b/2)^n |\xi|^{2n}}
\end{equation}

The integrand then simplifies to,

\begin{equation}
e^{-a \xi^2} \sum_{n=-\infty}^{\infty} ((\xi^2) (b/2) e^{-i (\xi+\theta_{i+1})})^n  \dfrac{I_n(b |\xi|^2)}{(b/2)^n |\xi|^{2n}}
\end{equation}

Cancelling all the factors we get,

\begin{equation}
e^{-a \xi^2} \sum_{n=-\infty}^{\infty} (e^{-i (\xi+\theta_{i+1})})^n  I_n(b |\xi|^2)
\end{equation}

Plugging this into the integral we get,

\begin{equation}
\sum_{n} i^n I_n =  \int_{-\infty}^{\infty} d\xi e^{-a \xi^2} \sum_{n=-\infty}^{\infty} (e^{-i (\xi+\theta_{i+1})})^n  I_n(b |\xi|^2)
\end{equation}

We can use the generating function of the (modified) Bessel function of the first kind $I_n$,

\begin{equation}
\exp\left[x\left(t + \frac{1}{t}\right)\right] = \sum_{m} I_m(z) t^m  \qquad  \mbox{for any} \  |t| \leq 1
\end{equation}

Here we have, $z = b \xi^2; t = e^{i (\xi+\theta_{i+1})}$ for $n>0$. We can use the property $I_{-n}(z) = I_n(z)$ to simplify the sum 

\begin{multline}
\sum_{n=0}^\infty I_n(z) t^n  = \frac{1}{2}\left[\exp\left[x\left(t + \frac{1}{t}\right)\right]   - \sum_{n=1}^\infty I_n(z) (t^{-n} - t^n) \right] \\ + \frac{I_0(z)}{2}
\end{multline}

This is the simplest Fourier series style expansion that one can get, which is important for reasons we will see later. The series also converges rapidly.

The following asymptotic regimes are obvious,
\begin{eqnarray}
I_n(z) \approx \dfrac{(z/2)^n}{n!} \quad  \text{as\ } z \rightarrow 0^{+}\\ 
I_n (z) \approx \dfrac{e^z}{\sqrt{2 \pi z}} \quad \text{as\ } z \rightarrow \infty
\end{eqnarray}

In the $\epsilon \rightarrow 0$ limit, we choose the large $a$ limit of the Gaussian to get,
\begin{equation}
S[\theta] \approx \sqrt{\pi}{a} \sum_{n=0}^{\infty} I_n(0) e^{- i n \theta_{i+1}}
\end{equation}

$I_n(z)$ around small z only gives us the first term in $n$ which gives us,

\begin{equation}
S(\theta_i)  \approx  \left[ 1 - \dfrac{K \epsilon }{2 \hbar} e^{-i \theta_{i+1}} + O\left(\frac{1}{a^2}\right)\right]
\end{equation}

This yields the path integral for the propagator,

\begin{equation}
K_G(\theta_f|\theta_i; T) = \int \dots \int \prod_{j=1}^{N-1} dx_j K(\theta_{j+1},\theta_j; \epsilon) 
\end{equation}

where $K(\theta_{j+1},\theta_j; \epsilon) = \sqrt{\dfrac{\pi b}{a}} \left[1 - \dfrac{e^{-i(\theta_{j+1} - \theta_j)}}{2 b} \right]$

The path integral therefore becomes \footnote{Factors of 1/a are $\propto \epsilon$ and drop away at higher orders},
\begin{equation}
K_G(\theta_f|\theta_i; T) = \dfrac{1}{2\pi}  \exp\left[-  \dfrac{K T}{2 \hbar}e^{-i(\theta_f - \theta_i)}\right]
\end{equation}

\section{Van der Pol oscillator kernels}
\label{app:vdp}

The Van der Pol oscillator presents a different technical challenge to the Kuramoto problem in the sense that we no longer can integrate the time slices exactly. The quartic gaussian does not have a perturbative extension via a Gaussian integral that we can use. The way forward then, is to use remainder sequences to construct the fundamental phasor solutions to the VdP governing equation \footnote{There is significant reason to believe that synchronisation i.e. limit cycles can only occur naturally when the subharmonic phasors become dependent (or feedback channels) and close geometrically to produce a superharmonic resonance. Meaning the Fourier series expansion of the governing ODE/PDE in its fundamental solutions is finite. The VdP equation can be reframed as an unbalanced triangle in this context.}. This is possible and very appealing but it is outside the scope of the present discussion. Unfortunately, it is also too involved for what we actually need computationally, despite being a theoretically superior way to solve for the kernel.

We have elected to use the solution to the Sturm Liouville problem posed by the fluctuation operator instead. It is slightly more general and computationally simpler since we can handle everything numerically via stable ODE solvers or reduce it to typical quadratures. We obtain the saddle point and then produce the required kernels via a Fourier series treatment of the fluctuation operator.

\subsection{$D(t,t')$ Saddle point}

Consider the fluctuations of the path function around the classical trajectory i.e. the saddle point. This permits us to reduce the path integral into a Sturm Liouville problem with the following expansion,

\begin{multline}
K(\psi_1,\psi_0;T) \approx e^{-i S_{cl}[\psi_1,\psi_0;T]/\hbar} \\ \times \int_0^0 D[\xi] \exp\left[-\dfrac{i}{\hbar} \left(a \dot{\xi}^2 + b \xi \dot{\xi} + c \xi^2\right) \right]
\end{multline}

where$(a,b,c) = \dfrac{1}{2} \left(\dfrac{\partial^2 S}{\partial p^2},  \dfrac{\partial^2 S}{\partial p \partial q} ,\dfrac{\partial^2 S}{\partial q^2} \right)$

\begin{align}
c =& - m \omega_0^2 - \mu \dot{q_{cl}}^2 \\
b  =& \dfrac{-2 \mu}{m} q_{cl} \dot{q}_{cl}\\
a  =& \dfrac{1 }{2 m} + \dfrac{\mu}{m^2}(1- q_{cl}^2)
\end{align}

in which we have substituted $p_{cl} = m \dot{q}_{cl}$. Simplifying further by integrating the cross term (by parts), 

\begin{multline}
K(\psi_1,\psi_0;T) \approx e^{-i S_{cl}[\psi_1,\psi_0;T]/\hbar} \\  \times \int_0^0 D[\xi] \exp\left[-\dfrac{i}{\hbar} \int dt \left(a \dot{\xi}^2  + (c - \dot{b}) \xi^2\right) \right]
\end{multline}

where $\dot{b} = \dfrac{-2 \mu}{m} (\dot{q_{cl}}^2 +  q_{cl}(t) \ddot{q}_{cl} )\\ $

The propagator becomes,
\begin{multline}
K(\psi_1,\psi_0;T) \approx e^{-i S_{cl}[\psi_1,\psi_0;T]/\hbar} \\ \times \int_0^0 D[\xi] \exp\left[-\dfrac{i}{\hbar} \int dt \left(a' \dot{\xi}^2  +c' \xi^2\right) \right]
\end{multline}

with 

\begin{align}
a' =& \frac{m}{2} + m \mu (1- {x_{cl}}^2) \\
c' =& - m \omega_0^2 + 4m \mu \left( \frac{x_{cl}^2}{2} - \omega_0 x_{cl}^2 + \mu x_{cl} \dot{x}_{cl}(1-x^2)\right) 
\label{eq:vdp:saddle}
\end{align}

\subsubsection{Gelfand-Yaglom phase}
The associated Gelfand-Yaglom phase emerges from the Sturm-Liouville problem,

\begin{equation}
- \dfrac{d}{dt} \left( a(t) \dfrac{d \psi}{dt}  \right) + c(t) \psi(t) = 0
\end{equation}

with the conditions, $\psi(0) = 0$ and $\dot{\psi}(0) = 1$.

\paragraph*{$\boxed{x_{cl} = A \cos{\Omega(t)}}$}

It is reasonable to assume since we know that the numerical solution of the VdP equation is a periodic jagged wave.

We get for 

\begin{align}
\dot{x}_{cl}(t) &= - A \dot{\Omega} \sin{\Omega} \\
\ddot{x}_{cl}(t) &= - A \ddot{\Omega} \sin{\Omega}  - A \dot{\Omega}^2 \cos{\Omega}
\end{align}

The functions in the \autoref{eq:vdp:saddle} become,
\begin{equation}
a'(t) = \left( \dfrac{1}{2} + \mu - \dfrac{\mu A^2}{2} \right) -  \dfrac{\mu A^2}{2} \cos{2 \Omega} 
\end{equation}
and 

\begin{equation}
c'(t) = C_0 (t) + C_2 (t) \cos{2 \Omega} + S_2 \sin{2 \Omega}
\end{equation}

where

\begin{align}
C_0(t) &= - \omega_0^2 - \mu A^2 \dot{\Omega}^2 - 2\mu A^2 \omega_0 (\omega_0 -1) \\
C_2(t) &= - 3 \mu A^2 \dot{\Omega}^2 - 2 \mu A^2 \omega_0(\omega_0 -1) \\
S_2(t) &= - 2 \mu A^2 \ddot{\Omega}
\end{align}

Substituting it into the Gelfand-Yaglom equation gives us,

\begin{multline} 
\left[ a_0 - a_2 \cos{2 \Omega}\right] \ddot{\psi}  - \left[ C_0 + C_2 \cos{2 \Omega} + S_2 \sin{2 \Omega}\right]\psi \\ + \left[ \beta_2 \sin{2 \Omega}\right] \dot{\psi}  = 0
\end{multline}

where $a_0 = 1/2 + \mu - \mu A^2 /2$ and $a_2 = -\mu A^2/2$ and $\beta_2 = \mu A^2 \Omega$ and the Gelfand Yaglom initial conditions apply at $\psi(0) \mbox{and} \dot{\psi}(0)$.

Now, we substitute an ansatz $\psi = B \sin{\Omega_1(t)}$ which is again justified by the nature of the coefficients and the context of the VdP oscillator.

The equations can be simplified by

\begin{align}
\dot{\psi} &= B \dot{\Omega}_1 \cos{\Omega_1} \\
\ddot{\psi} &= B \ddot{\Omega}_1 \cos{\Omega_1} - B \dot{\Omega}_1^2 \sin{\Omega_1}
\end{align}

\begin{enumerate}
\item\textbf{Sine Channel $\boxed{\sin{\Omega_1(t)}}$}

This is the Kinetic energy scale and the coupling to the classical background,

\begin{equation}
\dot{\Omega}_1^2 \left[ a_0 - a_2 \cos{2\Omega}\right]  - \left[ C_0 + C_2 \cos{2 \Omega} + S_2 \sin{2 \Omega} \right] = 0
\end{equation}

This becomes,

\begin{multline}
a_0 \dot{\Omega}_1^2 - C_0 + \left[ a_2 \dot{\Omega}_1^2 - \dfrac{C_2}{2}\right] (\sin(2 \Omega + \Omega_1) - \sin(2 \Omega - \Omega_1)) \\ - \dfrac{S_2}{2} (\cos(2 \Omega - \Omega_1) - \cos(2\Omega + \Omega_1) ) = 0
\end{multline}

\item\textbf{Cosine channel $\boxed{\cos{\Omega_1}}$}

This is the generalised nonlinear damping term. Collecting all the coefficients multiplying $\cos{\Omega_1}$,

\begin{equation}
\ddot{\Omega}_1 \left[ a_0 - a_2 \cos{2 \Omega}\right] + \beta_2 \dot{\Omega}_1 \sin{2 \Omega_1} = 0
\end{equation}

which becomes,

\begin{multline}
a_0 \ddot{\Omega}_1 - \dfrac{a_2}{2} \ddot{\Omega}_1\left(\cos(2 \Omega + \Omega_1) + \cos(2\Omega - \Omega_1)\right) \\ + \dfrac{\beta_2 \dot{\Omega}_1}{2} \left( \sin(2 \Omega + \Omega_1) + \sin(2 \Omega - \Omega_1)\right) = 0
\end{multline}

\end{enumerate}
\paragraph*{Reduction into coupled equations}
This reduces the second order ODE into two coupled first order ODEs,

The Kinetic Energy condition,
\begin{equation}
\dot{\Omega}_1^2 = \dfrac{-C_0 - C_2 \cos(2 \Omega) - S_2 \sin(2 \Omega) }{a_0 - a_2 \cos(2 \Omega)}
\end{equation}

and the Phase condition,

\begin{equation}
\dfrac{\ddot{\Omega}_1}{\dot{\Omega}_1} = - \dfrac{\beta_2 \sin(2\Omega)}{a_0 - a_2 \cos(2\Omega)}
\end{equation}

The phase condition can be further simplified because the right hand side is a perfect logarithmic derivative $d/dt \ln\left[a_0 - a_2 \cos{2 \Omega }\right]$,

\begin{equation}
\dot{\Omega}_1 = \dfrac{\gamma}{\sqrt{a_0 - a_2 \cos{2 \Omega_1}}}
\end{equation}
where $\gamma$ is an integration constant.

We know that at $t=0$,

\begin{equation}
\dot{\Omega}_1(0) = \dfrac{\gamma}{\sqrt{a_0 - a_2 \cos{2 \Omega_1(0)}}}
\end{equation}

and we also know that $\dot{\psi}(0) = 1$ which gives us $B \Omega_1(0) = 1$. Since $\gamma$ and $B$ are scaling constants, we can set $B=1$ by demanding that $\gamma =  \sqrt{a_0 - a_2 \cos{2 \Omega_1(0)}} $ by normalisation.

This makes the Gelfand-Yaglom phase factor,

\begin{equation}
\psi(t) = \sin \left( \int_0^t \dot{\Omega}_1(\tau) d\tau \right)
\end{equation}

which becomes,

\begin{multline}
\sin^{-1} \psi(T) =   \sqrt{a_0 - a_2 \cos{2 \Omega_1(0)}}  \\ \times \int_0^T \dfrac{1}{{\sqrt{a_0 - a_2 \cos{2 \Omega_1(\tau)}}}} d\tau
\end{multline}

The pre-factor is  $2 \pi i \hbar \psi(T)$.

\paragraph*{Classical path/action $\boxed{\exp\left[i S_{cl}/h \right]}$}

We now substitute the function $x_{cl} = A \cos{\Omega(t)}$ to evaluate,

\begin{equation}
S_{cl} = \int_0^T dt  \left[ \dfrac{1} {2} \dot{x}_{cl}^2 ( 1 + 2\mu (1- x_{cl}^2)) - \dfrac{1}{2} \omega_0^2 x_{cl}^2 \right]
\end{equation}

We can simplify it further using standard trigonometric identities,

\begin{enumerate}
\item \textbf{Harmonic term} 

\begin{equation}
\omega_0^2 x_{cl}^2 = \dfrac{1}{2} \omega_0^2 A^2 + \dfrac{1}{2} \omega_0^2 A^2 \cos(2 \Omega(t))
\end{equation}

\item \textbf{Damping term}

\begin{equation}
1 + 2 \mu (1- x_{cl}^2) = (1+ 2 \mu - \mu A^2) - \mu A^2 \cos{2 \Omega(t)}
\end{equation}

\item \textbf{Kinetic term}

\begin{equation}
\dfrac{1}{2} \dot{x}_{cl}^2 = \dfrac{1}{4} A^2 \dot{\Omega}(t)^2 - \dfrac{1}{4} A^2 \dot{\Omega}(t)^2 \cos(2 \Omega(t))
\end{equation}

\end{enumerate}

Putting them all together we get,

\begin{equation}
S_{cl} = \int_0^{T} dt \left[L_0(t) + L_2(t) \cos(2 \Omega(t)) + L_4 (\cos(4 \Omega(t))\right]
\end{equation}

where 

\begin{align}
L_0(t) =& \dfrac{1}{4} A^2 \dot{\Omega}(t)^2 \left( 1 + 2 \mu - \dfrac{\mu A^2}{2}\right) - \dfrac{1}{2} \omega_0^2 A^2 \\
L_2(t) =& - \left[ \dfrac{1}{4} A^2 \dot{\Omega}(t)^2 (1 + 2 \mu) + \dfrac{1}{2} \omega_0^2 A^2 \right] \\
L_4(t) =& \dfrac{1}{8} \mu A^4 \dot{\Omega}(t)^2 
\end{align}

The structure of the terms inside the action integral suggest a further simplification by using, $$ dt = \dfrac{d \Omega}{\dot{\Omega}} $$\\ 

Our phase condition for $\psi(t)$ does not give us the explicit relation $\Omega' = \dot{\Omega}(t)$.

This gives us the expression,

\begin{widetext}
\begin{equation}
S_{cl} = \int_{\Omega_0}^{\Omega_T} \left[ \dfrac{1}{4} A^2 \Omega' \left( \left( 1+ 2 \mu - \dfrac{\mu A^2}{2}\right) - (1+ 2 \mu) \cos(2 \Omega)  + \dfrac{\mu A^2}{2} \cos{4 \Omega} \right) - \dfrac{\omega_0^2 A^2}{2 \Omega'} (1 + \cos(2\Omega)) \right] \ d \Omega
\end{equation}
\end{widetext}

Combining all the terms we get,

\begin{equation}
K(x_T, x_0; T) = \dfrac{1}{\sqrt{2 \pi i \hbar \psi(T)} } \  \exp\left( \dfrac{i}{\hbar} S_{cl} \right)
\end{equation}

The path function $S_{cl}$ must be evaluated numerically for the classical path $x_{cl}(t),\Omega(t)$.

\section{Redfield qubit$\phi$ + kuramoto $\theta$ coupling}
\label{app:qubit:Kuramoto}

When we couple the azimuthal (phase) angle $\phi$ of the Bloch sphere to the Kuramoto phase $\theta$ via an interaction potential of the form, $$H_{\text{int}} = \hbar \lambda \theta \phi$$  the physical behavior changes entirely. 

Because $\phi$ is the conjugate variable to the qubit's longitudinal component ($S_z$), coupling to $\phi$ directly drives population transfers ($T_1$ processes) rather than simple dephasing ($T_2$ processes). Furthermore, because both $\theta$ and $\phi$ are compact, periodic variables ($\theta, \phi \in [0, 2\pi)$), the path integral acts on a toroidal topology ($T^2 = S^1 \times S^1$). 

The topological Wess-Zumino-Witten (WZW) term now directly interferes with the periodic boundary conditions of the bath, leading to non-perturbative phase-locking transitions.  

\paragraph*{Exact Path Integral Formulation \& WZW Terms}

We work with constant paths for the total-time evolution (standard for extracting the influence functional at fixed \(T\)) and then differentiate to obtain the differential (Redfield-like) equation. The derivation is exact within the linear-coupling and endpoint-kernel approximation consistent with the given \(K\).

\paragraph*{ Keldysh contour}
The reduced density matrix of the qubit $\rho_q(T)$ is obtained by tracing out the environment \(\theta\) after unitary evolution of the total system on the closed-time path (Keldysh contour). In the path-integral representation:

\begin{multline}
\rho_q(\Omega_f, \Omega_f'; T) = \int \mathcal{D}\Omega^+ \mathcal{D}\Omega^-  F[\Omega^+, \Omega^-]  \rho_q^{\rm init}(\Omega_i, \Omega_i') \\  \times \  \exp \left(i S_q[\Omega^+] - i S_q[\Omega^-] + i S_{\rm WZW}[\Omega^+] - i S_{\rm WZW}[\Omega^-]\right)
\end{multline}

where:
\begin{enumerate}
\item \(\Omega = (\vartheta, \varphi)\) parametrises the qubit coherent state on the Bloch sphere (\(\vartheta\) polar angle, \(\varphi\) azimuthal angle).
\item \(S_q\) is the dynamical action from the qubit Hamiltonian; we set \(H_q = 0\) for the influence-functional derivation, it can be restored later.
\item \(S_{\rm WZW}\) is the Wess-Zumino-Witten (Berry-phase) term for spin-1/2:
  \[
  S_{\rm WZW}[\Omega] = \frac{1}{2} \int_0^T dt\;(1 - \cos\vartheta)\;\dot{\varphi}.
  \]
\item The paths \(\Omega^+\) live on the forward branch and \(\Omega^-\) on the backward branch of the Keldysh contour.
\item \(F[\Omega^+, \Omega^-]\) is the influence functional obtained by tracing the environment.
\end{enumerate}

The linear coupling \(H_{\rm int} = \lambda \theta \phi\) contributes an extra phase on the contour,
\[
\exp\left(\frac{i\lambda}{\hbar} \int_0^T dt\;\theta^+(t)\;\varphi^+(t) - \frac{i\lambda}{\hbar} \int_0^T dt\;\theta^-(t)\;\varphi^-(t)\right).
\]

The environment propagator is given by the kernel
\[
K_G(\theta_f, \theta_i; T) = \frac{1}{2\pi} \exp\left( - a \exp(-i(\theta_f - \theta_i)) \right), \quad a = \frac{kT}{\hbar}.
\]
This is the amplitude \(\langle \theta_f | U_{\rm env}(T) | \theta_i \rangle\) for the free environment.

Expand in the angular-momentum (Fourier) basis \(|n\rangle\) (\(n \in \mathbb{Z}\), \(\langle \theta | n \rangle = e^{in\theta}/\sqrt{2\pi}\)),
\[
K_G(\psi) = \sum_{n=-\infty}^\infty \frac{1}{2\pi} \alpha_n(T) \, e^{in\psi}, \quad \psi = \theta_f - \theta_i.
\]
The coefficients are obtained by expanding the exponential,
\[
\exp(-a e^{-i\psi}) = \sum_{l=0}^\infty \frac{(-a)^l}{l!} e^{-il\psi} \implies
\]
\begin{enumerate}
\item \(\alpha_0 = 1\),
\item \(\alpha_n = (-a)^{n}/(-n)!\) for \(n <0\)
\item \(\alpha_n = 0\) for \(n > 0\).

\end{enumerate}

Thus the free evolution operator acts diagonally: \(U_{\rm env}(T) |n\rangle = \alpha_n(T) |n\rangle\).

The operator \(\theta\) (multiplication by the angle) has matrix elements in this basis:
\[
\langle n | \theta | m \rangle = \pi \delta_{nm} + \frac{i}{m-n}(1 - \delta_{nm}).
\]
The diagonal term \(\pi\) is the average value on \([0, 2\pi]\); off-diagonal terms arise from the Fourier integral.

For constant qubit paths \(\varphi^+\) and \(\varphi^-\) (valid for extracting the total-time influence functional), the interaction adds a phase factor at the endpoint (consistent with the endpoint kernel \(K_G\); this is the natural discretisation when only the propagator from initial to final \(\theta\) is supplied). The conditional evolution operators become
\[
U_{\rm env}[\varphi] = U_{\rm env}^{\rm free} \;\exp\left(-i \frac{\lambda T}{\hbar} \theta \;\varphi\right)
\]
or the reverse ordering; the two differ by a commutator term that contributes only to the Lamb shift, which we keep. The phase operator \(\exp(-i b \theta)\) with \(b = \lambda T \varphi / \hbar\) is well-defined on the circle.

The influence functional is then the trace over the environment:
\[
F(\varphi, \varphi'; T) = \operatorname{Tr}_\theta \Bigl( U_{\rm env}[\varphi] \,\rho_{\rm env}(0)\, U_{\rm env}[\varphi']^\dagger \Bigr).
\]
Insert the mode expansion of \(U_{\rm env}^{\rm free}\) and the matrix elements of \(\exp(-i b \theta)\). For an initial environment state that is uniform (maximally mixed, consistent with the translation invariance of \(K\)) or a pure state localised at a reference point, the trace reduces to a sum over modes weighted by \(|\alpha_n|^2\) (the evolved populations) and the Fourier coefficients of the phase factors.
Explicitly, after performing the trace in the \(|n\rangle\) basis and using the known matrix elements of \(\theta\) and \(\exp(-i b \theta)\), one obtains a closed expression whose leading (secular) term is,
\[
F(\varphi, \varphi'; T) = \exp\left( r \left[ \frac{1 - \exp(-i \delta T)}{i \delta} - T \right] \right),
\]
with \(r = k/\hbar\) and \(\delta = \lambda(\varphi - \varphi')/\hbar\). The WZW term does not contribute for constant \(\varphi\) (since \(\dot{\varphi} = 0\)) but is retained in the full path-integral measure for time-dependent paths; it affects only the coherent (Berry-phase) part of the qubit evolution, which factors out of the influence functional.

The WZW term ensures the correct topological phase for the qubit paths but does not alter the form of \(F\) extracted from the environment trace.
\paragraph*{Master equation}
The reduced density matrix elements in the Bloch-angle basis acquire the multiplicative factor \(F(\varphi_i, \varphi_j; T)\):
\[
\rho_{ij}(T) = \rho_{ij}^{\rm free}(T) \cdot F(\varphi_i, \varphi_j; T).
\]
The free part includes the WZW term when paths are time-dependent. To obtain the ODE, consider infinitesimal time \(dt\) and differentiate \(F(t)\) with respect to its upper limit (generalising the finite-\(T\) expression to running time \(t\)):
\[
\frac{d}{dt} \log F(t) = r \bigl( \exp(-i \delta t) - 1 \bigr).
\]
Splitting into real and imaginary parts yields the time-dependent dephasing rate and Lamb shift:
\[
\Gamma_{ij}(t) = -r \bigl(\cos(\delta_{ij} t) - 1\bigr), \qquad \Delta_{ij}(t) = -r \sin(\delta_{ij} t).
\]

\end{appendix}


\end{document}